\documentclass{aastex62}

\usepackage{lineno}
\submitjournal{The Astrophysical Journal}

\shorttitle{Type Ia SN optical spectroscopy by the CSP}
\shortauthors{Morrell et al.}
\begin{document}

\title{Optical Spectroscopy of Type Ia Supernovae by the Carnegie Supernova Projects I \& II}

\correspondingauthor{Nidia Morrell}
\email{nmorrell@lco.cl}

\author[0000-0003-2535-3091]{N.~Morrell}
\affil{Carnegie Observatories, Las Campanas Observatory, Casilla 601, La Serena, Chile}

\author[0000-0003-2734-0796]{M.~M.~Phillips}
\affil{Carnegie Observatories, Las Campanas Observatory, Casilla 601, La Serena, Chile}

\author[0000-0001-5247-1486]{G.~Folatelli}
\affil{Instituto de Astrof\'{\i}sica de La Plata (IALP), CONICET,
Paseo del Bosque  S/N, 1900, Argentina}
\affil{Facultad de Ciencias Astron\'omicas y Geof\'{\i}sicas (FCAG), 
Universidad Nacional de La Plata (UNLP), Paseo del Bosque S/N,
1900, Argentina}

\author[0000-0002-5571-1833]{M.~D.~Stritzinger}
\affil{Department of Physics and Astronomy, Aarhus University, Ny Munkegade 120, DK-8000 Aarhus C, Denmark}

\author[0000-0001-7981-8320]{M.~Hamuy}
\affil{Fundaci\'on Chilena de Astronom\'{\i}a, Santiago, Chile}

\author[0000-0002-8102-181X]{N.~B.~Suntzeff}
\affil{George P. and Cynthia Woods Mitchell Institute for Fundamental Physics and Astronomy, 
Texas A\&M University, Department of Physics and Astronomy, College Station, TX 77843, USA}

\author[0000-0003-1039-2928]{E.~Y.~Hsiao}
\affil{Department of Physics, Florida State University, 77 Chieftan Way, Tallahassee, FL 32306, USA}

\author[0000-0002-2387-6801]{F.~Taddia}
\affil{Department of Physics and Astronomy, Aarhus University, Ny Munkegade 120, DK-8000 Aarhus C, Denmark}

\author[0000-0003-4625-6629]{C.~R.~Burns}
\affil{Observatories of the Carnegie Institution for Science, 813 Santa Barbara St, Pasadena, CA 91101, USA}

\author[0000-0002-4338-6586]{P.~Hoeflich}
\affil{Department of Physics, Florida State University, 77 Chieftan Way, Tallahassee, FL 32306, USA}

\author[0000-0002-5221-7557]{C.~Ashall}
\affil{Department of Physics, Virginia Polytechnic Institute and State University, 850 West Campus Drive,
Blacksburg, VA 24061, USA}

\author[0000-0001-6293-9062]{C.~Contreras}
\affil{Carnegie Observatories, Las Campanas Observatory, Casilla 601, La Serena, Chile}

\author[0000-0002-1296-6887]{L.~Galbany}
\affil{Institute of Space Sciences (ICE, CSIC), Campus UAB, Carrer de Can Magrans, s/n, E-08193 Barcelona, Spain}
\affil{Institut d'Estudis Espacials de Catalunya (IEEC), E-08034 Barcelona, Spain}

\author[0000-0002-3900-1452]{J.~Lu}
\affil{Department of Physics, Florida State University, 77 Chieftan Way, Tallahassee, FL 32306, USA}

\author[0000-0001-6806-0673]{A.~L.~Piro}
\affiliation{Observatories of the Carnegie Institution for Science, 813 Santa Barbara St, Pasadena, CA 91101, USA}

\author{J.~Anais}
\affiliation{Carnegie Observatories, Las Campanas Observatory, Casilla 601, La Serena, Chile}

\author[0000-0001-5393-1608]{E.~Baron}
\affiliation{Planetary Science Institute, 1700 East Fort Lowell Road, Suite 106, Tucson, AZ 85719-2395, USA}
\affiliation{Hamburger Sternwarte, Gojenbergsweg 112, D-21029 Hamburg, Germany}
\affiliation{Dept of Physics \& Astronomy, University of Oklahoma, Norman, OK 73019, USA}

\author[0000-0002-5380-0816]{A.~Burrow}
\affiliation{Dept of Physics \& Astronomy, University of Oklahoma, Norman, OK 73019, USA}

\author[0000-0001-9952-0652]{L.~Busta}
\affiliation{Carnegie Observatories, Las Campanas Observatory, Casilla 601, La Serena, Chile}

\author{A.~Campillay}
\affiliation{Carnegie Observatories, Las Campanas Observatory, Casilla 601, La Serena, Chile}
\affiliation{Departamento de F\'{i}sica, Universidad de La Serena, Cisternas 1200, La Serena, Chile}

\author{S.~Castell\'{o}n}
\affiliation{Carnegie Observatories, Las Campanas Observatory, Casilla 601, La Serena, Chile}

\author{C.~Corco}
\affiliation{Carnegie Observatories, Las Campanas Observatory, Casilla 601, La Serena, Chile}
\affiliation{SOAR Telescope, Casilla 603, La Serena, Chile}

\author[0000-0002-0805-1908]{T.~Diamond}
\affiliation{Department of Physics, Florida State University, 77 Chieftan Way, Tallahassee, FL  32306, USA}
\affiliation{Laboratory of Observational Cosmology, Code 665, NASA Goddard Space Flight Center, Greenbelt, MD 20771, USA}

\author[0000-0003-3431-9135]{W.~L.~Freedman}
\affil{Department of Astronomy \& Astrophysics \& Kavli Institute for Cosmological Physics, University of Chicago, 
5640 South Ellis Avenue, Chicago, IL 60637, USA}

\author{C.~Gonzalez}
\affiliation{Carnegie Observatories, Las Campanas Observatory, Casilla 601, La Serena, Chile}

\author[0000-0002-6650-694X]{K.~Krisciunas}
\affil{George P. and Cynthia Woods Mitchell Institute for Fundamental Physics and Astronomy, Texas
A\&M University, Department of Physics and Astronomy, College Station, TX 77843, USA}

\author[0000-0001-8367-7591]{S.~Kumar}
\affil{Department of Physics, Florida State University, 77 Chieftan Way, Tallahassee, FL 32306, USA}

\author[0000-0003-0554-7083]{S.~E.~Persson}
\affil{Observatories of the Carnegie Institution for Science, 813 Santa Barbara St, Pasadena, CA 91101, USA}

\author[0000-0002-8303-776X]{J.~Ser\'{o}n}
\affiliation{Cerro Tololo Inter-American Observatory/NSF's NOIRLab, Casilla 603, La Serena, Chile}

\author[0000-0002-9301-5302]{M.~Shahbandeh}
\affil{Department of Physics, Florida State University, 77 Chieftan Way, Tallahassee, FL 32306, USA}]

\author{S.~Torres}
\affiliation{SOAR Telescope, Casilla 603, La Serena, Chile}

\author[0000-0002-9413-4186]{S.~A.~Uddin}
\affil{Center for Space Studies, American Public University System, 111 W. Congress Street, Charles Town, WV 25414, USA}

\author[0000-0003-0227-3451]{J.~P.~Anderson}
\affil{European Southern Observatory, Alonso de C\'ordova 3107, Casilla 19, Santiago, Chile}
\affil{Millennium Institute of Astrophysics MAS, Nuncio Monsenor Sotero Sanz 100, Off.104, Providencia, Santiago, Chile}

\author[0000-0003-0424-8719]{C.~Baltay}
\affiliation{Department of Physics, Yale University, 217 Prospect Street, New Haven, CT 06511, USA}

\author[0000-0002-8526-3963]{C.~Gall}
\affiliation{Department of Physics and Astronomy, Aarhus University, Ny Munkegade 120, DK-8000 Aarhus C, Denmark}
\affiliation{DARK, Niels Bohr Institute, University of Copenhagen, Jagtvej 128, 2200 Copenhagen, Denmark}

\author[0000-0002-4163-4996]{A.~Goobar}
\affiliation{The Oskar Klein Centre, Department of Physics, Stockholm University, SE-106 91 Stockholm, Sweden}

\author{E.~Hadjiyska}
\affiliation{Department of Physics, Yale University, 217 Prospect Street, New Haven, CT 06511, USA}

\author[0000-0002-3415-322X]{S.~Holmbo}
\affiliation{Department of Physics and Astronomy, Aarhus University, Ny Munkegade 120, DK-8000 Aarhus C, Denmark}

\author[0000-0002-5619-4938]{M.~Kasliwal}
\affiliation{Caltech, 1200 East California Boulevard, MC 249-17, Pasadena, CA 91125, USA}

\author[0000-0003-1731-0497]{C.~Lidman}
\affiliation{The Research School of Astronomy and Astrophysics, Australian National University, ACT 2601, Australia}

\author[0000-0002-2966-3508]{G.~H.~Marion}
\affiliation{Department of Astronomy, University of Texas at Austin, 2515 Speedway Stop C1400, Austin, TX, 78712-1205, USA}

\author[0000-0001-6876-8284]{P.~A.~Mazzali}
\affiliation{Astrophysics Research Institute, Liverpool John Moores University, IC2, Liverpool Science Park, 146 Brownlow Hill, Liverpool L3 5RF, UK}
\affiliation{Max-Plank-Institut f\"ur Astrophysik, Karl-Schwarszchild Str. 1, D-85748 Garching, Germany}

\author[0000-0002-3389-0586]{P.~Nugent}
\affil{Physics Department, University of California, Berkeley, CA 94720, USA}
\affil{Lawrence Berkeley National Laboratory, Department of Physics, 1 Cyclotron Road, Berkeley, CA 94720, USA}

\author[0000-0002-4436-4661]{S.~Perlmutter}
\affiliation{Lawrence Berkeley National Laboratory, Department of Physics, 1 Cyclotron Road, Berkeley, CA 94720, USA}
\affiliation{Physics Department, University of California, Berkeley, CA 94720, USA}

\author[0000-0003-0006-0188]{G.~Pignata}
\affiliation{Instituto de Alta Investigación, Universidad de Tarapacá, Casilla 7D, Arica, Chile}  

\author[0000-0003-4961-7653]{D.~Rabinowitz}
\affiliation{Department of Physics, Yale University, 217 Prospect Street, New Haven, CT 06511, USA}

\author{M.~Roth},
\affiliation{Carnegie Observatories, Las Campanas Observatory, Casilla 601, La Serena, Chile}
\affiliation{GMTO Corporation, Presidente Riesco 5335, Of. 501, Nueva Las Condes, Santiago}

\author[0000-0003-4501-8100]{S.~D.~Ryder}
\affiliation{School of Mathematical and Physical Sciences, Macquarie University, NSW 2109, Australia}
\affiliation{Astrophysics and Space Technologies Research Centre, Macquarie University, Sydney, NSW 2109, Australia}

\author[0000-0003-4631-1149]{B.~J.~Shappee}
\affiliation{Institute for Astronomy, University of Hawaii, 2680 Woodlawn Drive, Honolulu, HI 96822, USA}

\author[0000-0001-8764-7832]{J.~Vink\'o}
\affiliation{Konkoly Observatory, CSFK, MTA Centre of Excellence, Konkoly Thege M. \'ut 15-17, Budapest, 1121, Hungary}
\affiliation{ELTE E\"otv\"os Lor\'and University, Institute of Physics and Astronomy, P\'azm\'any P\'eter s\'et\'any 1/A, Budapest, 1117 Hungary}
\affiliation{Department of Experimental Physics, University of Szeged, D\'om t\'er 9, Szeged, 6720, Hungary}
\affiliation{Department of Astronomy, University of Texas at Austin, 2515 Speedway Stop C1400, Austin, TX, 78712-1205, USA}

\author[0000-0003-1349-6538]{J.~C.~Wheeler}
\affiliation{Department of Astronomy, University of Texas at Austin, 2515 Speedway Stop C1400, Austin, TX, 78712-1205, USA}

\author[0000-0001-6069-1139]{T.~de~Jaeger}
\affiliation{LPNHE, CNRS/IN2P3 \& Sorbonne Universit\'e, 4 place Jussieu, 75005 Paris, France}

\author[0000-0003-1523-9164]{P.~Lira}
\affiliation{Departamento de Astronom\'{\i}a, Universidad de Chile, Camino del Observatorio 1515, Santiago, Chile}

\author[0000-0002-6799-1537]{M.~T.~Ruiz}
\affiliation{Departamento de Astronom\'{\i}a, Universidad de Chile, Camino del Observatorio 1515, Santiago, Chile}

\author[0000-0002-5807-5078]{J.~A.~Rich}
\affiliation{Observatories of the Carnegie Institution for Science, 813 Santa Barbara St, Pasadena, CA 91101, USA}

\author[0000-0003-1072-2712]{J.~L.~Prieto}
\affiliation{Instituto de Estudios Astrof\'isicos, Facultad de Ingenier\'ia y Ciencias, Universidad Diego Portales, Avenida Ej\'ercito Libertador 441, Santiago, Chile }

\author[0000-0003-0483-5083]{F.~Di~Mille}
\affil{Carnegie Observatories, Las Campanas Observatory, Casilla 601, La Serena, Chile}

\author[0000-0003-0412-9664]{D.~Osip}
\affil{Carnegie Observatories, Las Campanas Observatory, Casilla 601, La Serena, Chile}

\author[0000-0003-4218-3944]{G.~Blanc}
\affil{Carnegie Observatories, Las Campanas Observatory, Casilla 601, La Serena, Chile}

\author{P.~Palunas}
\affil{Carnegie Observatories, Las Campanas Observatory, Casilla 601, La Serena, Chile}

\begin{abstract}
We present the second and final release of optical spectroscopy of Type~Ia Supernovae (SNe~Ia) obtained during the first and second phases of the \textit{Carnegie Supernova Project} (CSP-I and CSP-II). The newly released data consist of 148 spectra of 30 SNe~Ia observed in the course of the CSP-I, and 234 spectra of 127 SNe~Ia obtained during the CSP-II. We also present 216 optical spectra of 46 historical SNe~Ia, including 53 spectra of 30 SNe~Ia observed by the Cal\'an/Tololo Supernova Survey.  We combine these observations with previously published CSP data and publicly-available spectra to compile a large sample of measurements of spectroscopic parameters at maximum light, consisting of pseudo-equivalent widths and expansion velocities of selected features, for 232 CSP and historical SNe~Ia (including more than 1000 spectra).  Finally, we review some of the strongest correlations between spectroscopic and photometric properties of SNe~Ia. Specifically, we define two samples: one consisting of SNe Ia discovered by targeted searches (most of them CSP-I objects) and the other composed of SNe Ia discovered by untargeted searches, which includes most of the CSP-II objects. The analysed correlations are similar for both samples. We find a larger incidence of SNe Ia belonging to the Cool (CL)and Broad Line (BL) Branch subtypes among the events discovered by targeted searches, Shallow Silicon (SS) SNe~Ia are present with similar frequencies in both samples, while Core Normal (CN) SNe Ia are more frequent in untargeted searches.
\end{abstract}


\section{Introduction} \label{sec:intro}
The number of observational studies of Type~Ia supernovae (SNe~Ia) has continually increased as the result
of their great importance to multiple fields of astrophysics.  For example, SNe~Ia are the primary
source of iron-peak elements in the Universe, and their energy input plays an important role in the heating of
 interstellar matter in galaxies.  They also provide important constraints on binary evolution in the
Galaxy. But, probably their most outstanding property is that they are excellent extragalactic distance indicators,
e.g. \citet{zwicky1961},
and thus powerful tools for the determination of the cosmological expansion rate as a function of look-back time
(e.g. \citealt{burns2018, freedman2021, khetan2021, riess2022, uddin2023}). 

The \textit{Carnegie Supernova Project} began taking data in 2004 with the expressed goal of 
obtaining high-precision optical and near-infrared photometry of a large sample of SNe~Ia acquired in
well-understood photometric systems in order to characterize the ultimate precision of these events
for determining distances \citep{hamuy06}.  During its first phase (CSP-I), which operated for five
years between 2004-2009, light curves were obtained for 123 nearby SNe~Ia \citep{contreras10,stritzinger11,krisciunas2017}.
A large number of optical spectra were also acquired for these SNe~Ia, 604 of which were published by \citet{folatelli2013}.  
During a second phase of the project (CSP-II), carried out between 2011-2015, 
optical and near-infrared photometry were obtained for 214 SNe~Ia, 125 of which were located in the smooth 
Hubble flow at redshifts $0.027 < z < 0.137$ \citep{phillips19}.  A major goal of the CSP-II was
to obtain near-infrared spectroscopy, and more than 650 such spectra were acquired of 157 SNe~Ia
\citep{hsiao19}.  A significant number of optical spectra were also obtained, many for classification purposes, 
while more extensive follow-up was  performed for a limited number of events.

In this data release paper, we present 148 previously unpublished optical
spectra of 30 SNe Ia observed during the course of the CSP-I and 234 previously unpublished spectra of 127 SNe~Ia observed during the CSP-II. 
The CSP-I spectra presented here were not included in \citet{folatelli2013} because at the time of publication of that paper, definitive photometry was not yet available for the corresponding objects and as a consequence, the analysis of spectroscopic and photometric properties could not be carried out in the same way as for the other SNe Ia.
CSP optical spectra have already been published for SN~2010ae \citep{stritzinger2014}, SN~2011iv \citep{gall2018}, SN~2012Z \citep{max2015},  SN~2012fr \citep{childress2013}, iPTF13ebh \citep{hsiao2015}, SN~2013gy \citep{holmbo2019},
ASASSN-14lp \citep{shappee16}, LSQ14fmg \citep{hsiao20}, ASASSN-15hy \citep{lu21},  SN~2015bp \citep{wyatt21}, SN~2007if, SN~2009dc, LSQ12gpw, SN~2013ao, CSS140501-170414+174838 and SN~2015M \citep{ashall2021}, and SN~2015bo \citep{hoogendam2021}.  
Also, a large number of CSP optical spectra of SNe~Ia obtained near maximum light have 
been analysed by \citet{burrow2020} using Gaussian mixture models. The approach in this paper differs from the latter in that we intend to derive spectroscopic parameters at the time of maximum light for the SNe~Ia in our sample, and only those objects for which such parameters were obtained are considered in the discussion that follows.
 
In addition, we include in this paper a number of spectra of what we shall refer to as ``historical'' SNe~Ia.  These consist of:

\begin{itemize}
\item 53 optical spectra of 30 SNe~Ia observed by the Cal\'an/Tololo Supernova Survey \citep{hamuy1993}.
\item 163 optical spectra of an additional 16 SNe~Ia obtained mostly by members of the Cal\'an/Tololo team.  Severa of these spectra have been included in previous publications: SN~1986G \citep{phillips1987}, SN~1989B \citep{wells1994}, SN~1991T \citep{phillips1992}, 
SN~1991bg \citep{leibundgut1993}, SN~1992A \citep{kirshner1993}, and SN~1992K \citep{hamuy1994}, but are not yet publicly available on WISeREP \citep{wiserep}. 
Some of the spectra were acquired during the execution of the
Supernova Optical and Infrared Survey (SOIRS, PI M. Hamuy, 1999-2000) \citep{hamuy2001,hamuy2002,hamuy2002_erratum}.
\end{itemize} 

The primary goal of this paper is to make these 598 optical spectra available to the community.  We combine them with the
604 CSP-I spectra published by \citet{folatelli2013} and the previously-published CSP-I and CSP-II optical spectra mentioned above to re-examine some of the correlations between spectral and photometric properties presented by these authors.  
Results for the SNe~Ia drawn from targeted searches are compared with those discovered in untargeted (``blind'') searches.
Incorporated in this new analysis is the color stretch parameter $s_{BV}$  \citep{burns14}, which is more effective at characterizing the fastest-declining SNe~Ia than the widely-used $\Delta{\rm m}_{15}(B)$
parameter \citep{phillips93}.
Note that all the newly-presented data, including spectra of CSP~I~\&~II targets and the historical SNe~Ia,
will be released electronically via WISeREP. 

The outline of this paper is as follows: in \S2 the different spectroscopic data sets are detailed;
in \S3 measurements of expansion velocities and pseudo-equivalent widths are presented along with notes on a few specific SNe;
in \S4 we re-examine some of the spectroscopic and photometric correlations derived by \citet{folatelli2013} using the measurements presented in this paper, augmented by those of the SNe~Ia published by these authors;
finally, in \S5 a brief summary of the results is given.

\section{Data} \label{sec:obs}

Summaries of the new observations, classifications and photometric
properties of the SNe~Ia considered in this paper are
presented in Table~\ref{CSP-prop} (for CSP-I and CSP-II
data)  and Table~\ref{CT-prop} (for the historical SNe~Ia).
Photometric parameters for the CSP~I~\&~II targets are taken from 
Uddin et al. (accepted for publication in ApJ).
For the historical SNe~Ia, 
 template light-curve fits were computed using \texttt{SNooPy} \citep{burns2011} to the available photometry of our objects retrieved from  the references given in the final column of Table~\ref{CT-prop}.

In successive columns of Tables~\ref{CSP-prop} and \ref{CT-prop} we list: the SN identification; the number of new spectra released in this work (three dots
mean that no new spectra are presented); the phase range
covered by them; the supernova classification obtained from running \texttt{SNID} \citep{blondin2007}
on the earliest spectrum available to us; the Wang \citep{wang2009} subtype; the Branch \citep{branch2006} subtype; the heliocentric redshift\footnote{The redshift quoted here is not precisely the redshift as defined
in cosmology, in that it can contain peculiar velocities due to galaxy infall. If an averaged peculiar velocity of 300 km~s$^{-1}$ is assumed, it would add a 0.001 uncertainty in the redshift, as estimated from the spectroscopic
velocity.} 
of the host as listed in \citet{krisciunas2017} and
\citet{phillips19} except for LSQ12hnr, discussed in Section~\ref{sec:notes}; 
the time of maximum light; and the decline rate
($\Delta{\rm m}_{15}$)\footnote{$\Delta{\rm m}_{15}$ is approximately
equivalent to $\Delta{\rm m}_{15}(B)$, but is measured via \texttt{SNooPy}  fits to all photometric filters available, rather than being a direct measurement of the $B$-band light-curve decline rate.} and color stretch ($s_{BV}$) parameters from
\texttt{SNooPy} fits (unless otherwise specified).

Journals of the spectroscopic observations,
including details about the telescopes and instruments used, are presented in
Tables \ref{CSP-spectra} and \ref{CT-spectra} for the CSP and historical SNe Ia, 
respectively.

The optical spectra obtained by the CSP collaboration were reduced using standard IRAF\footnote{IRAF was distributed by 
the National Optical Astronomy Observatory,
which is operated by the Association of Universities for Research
in Astronomy (AURA) under cooperative agreement with the National Science
Foundation.} routines as described in \citet{hamuy06}.
 Briefly, reductions included bias subtraction, flat fielding, wavelength
calibration with arc lamp exposures obtained right before or after the SN
observation, and flux
calibration with spectra of spectrophotometric standard stars observed during the same night as the science targets. 
In the vast majority of the cases, the slit was oriented according to the parallactic angle. A few exceptions were made when the
parallactic orientation would result in a bright host nucleus significantly contaminating the supernova spectrum.
At least one  telluric standard chosen from \citet{bessell1999} was also observed during each observing night with
the same slit width as that used for the SN observation in order to correct for telluric absorption features.
When a telluric standard spectrum was not obtained on the same night as the science observations, 
no telluric correction was attempted.  Such spectra are labeled as
not corrected for telluric absorption in Table~\ref{CSP-spectra}.

The spectra obtained with the GMOS instrument on Gemini were reduced
following standard procedures with the IRAF Gemini/{\sc GMOS} package.

Optical spectra taken with the Nordic Optical Telescope (NOT) were obtained using ALFOSC (Andalucia Faint Object Spectrograph and Camera) with
grism \#4. Data reduction of NOT spectra was performed following standard procedures using a set of custom \texttt{MatLab} scripts written by F. Taddia. Flux calibration was performed using sensitivity functions derived from observations of standard star(s) obtained on the same night as the science observations.

As in \citet{folatelli2013}, in the last column of Table~\ref{CSP-spectra}, we provide the root mean square (rms) of the differences between two or more synthetic magnitudes calculated from the spectra and observed magnitudes in the same filters interpolated to the exact time of the spectral observation from the CSP light curves, after removal of a constant flux term.  This number provides a
measure of the correctness of the {\em shape} of the flux-calibrated spectrum.
In some cases we were not able to perform this comparison, either due to the restricted wavelength coverage of the spectrum, or because the
photometric data did not cover the epoch of the spectrum.
When the rms was larger than $\sim$0.15~mag for
at least three bandpasses, we used a low-order polynomial
function to correct the overall shape of the spectrum.

Most of the spectra of the historical SNe~Ia were also calibrated using 
standard IRAF routines.  Unfortunately, for some of these observations, details of the exposure time, airmass, and, in a few cases, the instrument used have been lost.  No attempt has been made to quantify errors in the shapes of the spectra introduced in the flux calibration through comparison with available photometry. 

\section{Measurements} \label{sec:measurements}

\subsection{Expansion velocities and pseudo-equivalent widths}

We measured expansion velocities and pseudo-equivalent widths (pW) of
selected features as described in \citet{folatelli2013}.
Following the definitions in \citet{garavini2007}
\citep[see Figure 4 and Table 5 of][]{folatelli2013} 
we performed measurements of the following features: 
\ion{Ca}{2} H\&K (pW1), \ion{Si}{2} $\lambda$4130 (pW2),
 \ion{Mg}{2} $\lambda$4481 (pW3;
blended with \ion{Fe}{2} and dominated by \ion{Ti}{2} in the Branch CL
class), \ion{Fe}{2} at $\sim$4600 \AA\ 
(pW4, blended with \ion{S}{2}), the \ion{S}{2} ``W'' feature at 
$\sim$5400 \AA\ (pW5),
\ion{Si}{2} $\lambda$5972 (pW6), \ion{Si}{2} $\lambda$6355 (pW7), 
and the \ion{Ca}{2} ``IR triplet'' (pW8).

All measurements were made by means of the IRAF {\it splot} task 
from the {\it onedspec} package.
Line expansion velocities were derived from Gaussian fitting of the
minimum of each absorption trough.  The pW measurements were
obtained by direct integration between two defined pseudo-continuum  
positions. Error estimates were computed with {\it splot} setting the 
number of samples for error computation (the {\it nerrsample} parameter) 
to 100, the {\it sigma0} parameter (uniform component of the uncertainty) to the rms flux measured in the
nearby pseudo-continuum, and disregarding any Poissonian component
of the uncertainty (i.e. setting the {\it invgain} parameter to zero).
We adopted 1 \AA~as the minimum realistic uncertainty for pW measurements.
For some  critical features, such as \ion{Si}{2} $\lambda$5972,
when unable to obtain a reliable measurement because the feature was very 
weak or not detected, we estimated upper limits of pW 
 considering the signal-to-noise ratio in the spectral 
region where such a feature would be expected.

\subsection{Spectroscopic parameters at maximum light}

The measurements described above for our own spectra were analysed in combination
with similar measurements
of spectra of the targets in Table \ref{CSP-prop} that are publicly available through WISeREP, specifically when our observations were not
sufficiently close to maximum light to compute the desired spectroscopic parameters.  
The procedure followed is the same described in \citet{folatelli2013} and 
is summarized as follows:

\begin{itemize}

\item In cases where several spectra encompassing maximum light were available to us, we fit low-order polynomials to our pW and expansion velocity measurements and used those 
functions to interpolate the values at maximum.

\item When two spectra were available in the interval $-4$ to $+4$ days from maximum, we interpolated values at maximum from them.

\item If only data before or after maximum were available, but one spectrum was
obtained within 1 day of maximum light, an extrapolation was allowed.

\item In the most frequent
cases where only one spectrum was available within the range $-4$ to $+4$ days from maximum, we applied the slopes
given in Tables 4 and 7 of \citet{folatelli2013} to estimate
 values at maximum light, combining the errors estimated for our measurements with those
 coming from the assumed slopes.
 
 \end{itemize}

This procedure allowed determination of pWs and expansion velocities at maximum light for the selected features in 15 CSP~I   SNe~Ia not included in our previous spectroscopic release  \citep{folatelli2013},
113 SNe~Ia observed by the CSP~II, and 27 historical SNe~Ia.
All targets for which we were able to derive spectroscopic parameters at maximum
light are included, regardless of their being considered as ``normal'' or ``peculiar'' SNe Ia.
The values of the pWs at maximum light are presented in Table~\ref{pWs}. 
Measurements of expansion velocities at maximum
light can be found in Table~\ref{vels}. 
Note that the SNe are sorted by Branch type \citep{branch2006} in both tables.

\subsection{Notes on particular targets} \label{sec:notes}

{\bf LSQ12ca}: This SN~Ia has the lowest value of $\Delta{\rm m}_{15}$ 
in the CSP-II sample (0.618 $\pm$ 0.081 mag.), although
its $s_{BV}$ (1.195 $\pm$ 0.097), while high, is comparable to that of other normal SNe~Ia. From the spectrum available to us, obtained at phase 2.8 days past maximum,
 we derive normal values for its expansion velocities
and pseudo-equivalent widths at maximum light. \ion{C}{2} absorption
is probably present on the red side of the \ion{Si}{2} $\lambda$6355 line.

{\bf LSQ12gpw}: There are three public spectra of this SN~Ia from PESSTO\footnote{Public 
ESO Spectroscopic Survey of Transient Objects,
\citet{smartt15}.}, available in WISeREP: the one obtained at phase -1.5 days and
the other two at phase -0.5 days.
There is also  one CSP spectrum
at phase $+5.8$ days. All these spectra show an
absorption line, redward of \ion{Si}{2} $\lambda$6355, well
separated from the \ion{Si}{2} feature that, if identified as  C\,{\sc ii} $\lambda$6580, yields
expansion velocities of approximately 7,500 km\,s$^{-1}$, 7,400 km\,s$^{-1}$ and
6,200 km\,s$^{-1}$, for phases $-1.5$, $-0.5$, and $+5.8$ days, respectively.

{\bf LSQ12hno} shows expansion velocities somewhat lower than usual 
($\sim$ 8000-9000~km\,s$^{-1}$) in our two spectra obtained at 2.4 and 1.5 days before maximum light, respectively.

{\bf LSQ12hnr}: Two spectra of LSQ12hnr are available in WISeREP, both
obtained by PESSTO, at phases $+0.8$ and $+10.8$ {\bf days}, respectively.
The classification report by \citet{atel4673} gave a possible
redshift of $z = 0.135$ inferred from the SN~Ia spectrum.  No obvious host
is detected in our follow-up images or in a deep VLT-MUSE observation
of the SN~Ia site. 
However, an apparent cluster of galaxies is observed
whose brightest member lies about 40\arcsec~West of the SN~Ia location.
We determined redshifts for the three brightest galaxies in that cluster 
(from a total of at least four) obtaining a weighted average of 
$z = 0.1243 \pm 0.0002$, the value we
therefore adopted for LSQ12hnr assuming this SN~Ia occurred in a fainter
member of the same cluster of galaxies.

{\bf OGLE-2013-SN-015}: There is only one CSP spectrum of this
SN~Ia, which, albeit noisy, seems normal, and was obtained very close to maximum
light (phase $= +0.7$ days). However, our photometric follow-up was poor for this
target and consequently we decided not to consider this SN~Ia in our
analysis of SN properties at maximum light.

\begin{figure}
\begin{center}
\includegraphics[width=0.8\textwidth]{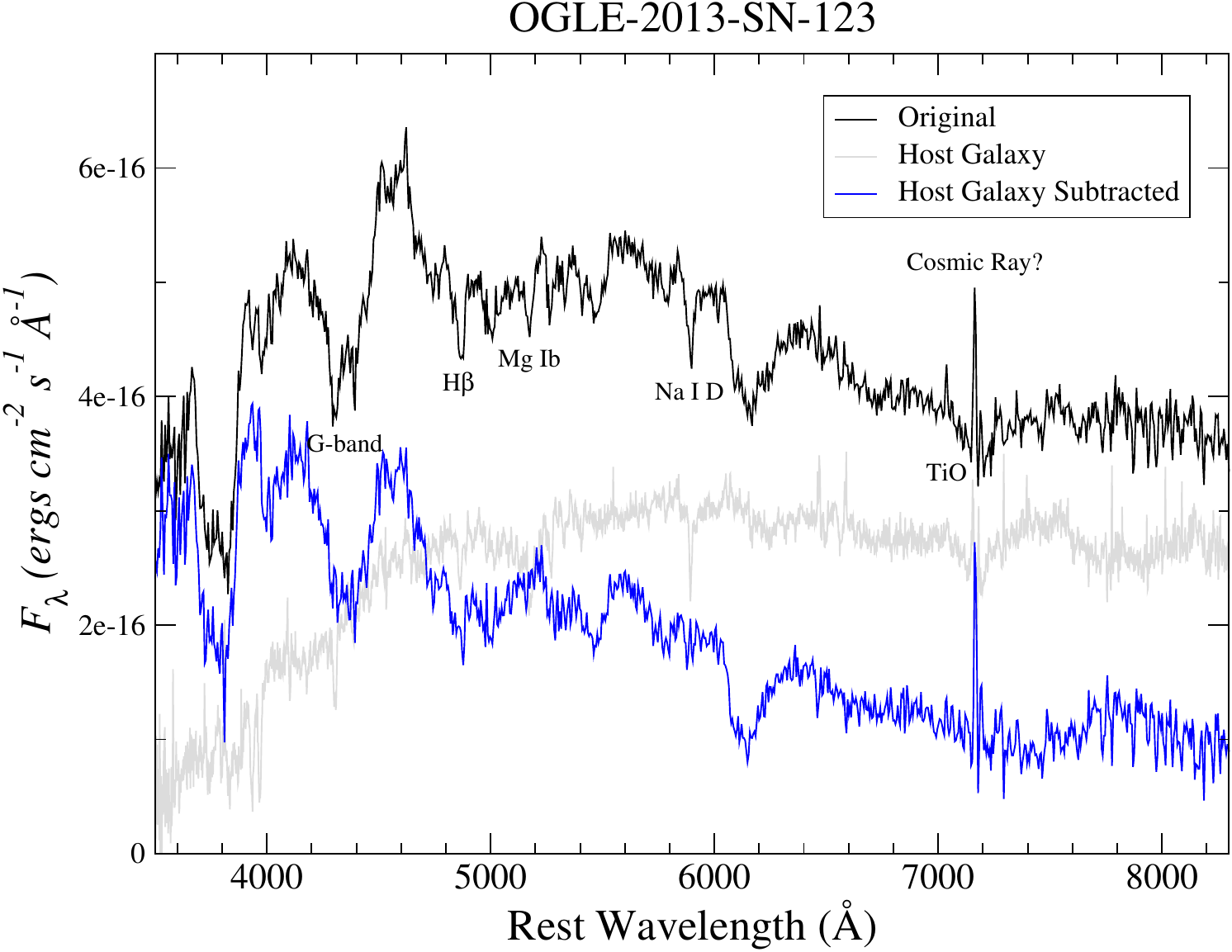}
\end{center}
\caption {The spectrum of OGLE-2013-SN-123 after galaxy subtraction. The black spectrum is
the original PESSTO observation, the gray spectrum is that of the host galaxy obtained by the
CSP-II, and the blue spectrum is the difference between the two after scaling the host-galaxy 
spectrum to minimize the stellar features in the original PESSTO spectrum.  Due to the differing
wavelength resolution of the spectra, some residual subtraction features are evident (e.g., for
the \ion{Na}{1}~D line). See text for further details.
\label{fig:CSP13aao}}
\end{figure}

{\bf OGLE-2013-SN-123}: There is only one spectrum available from
WISeREP, obtained by the PESSTO collaboration at maximum light,
which shows clear evidence of host galaxy contamination and, therefore, the pWs measured from it
are unreliable as well as the spectral type determined via SNID.  However, we
obtained a spectrum of the host galaxy with the WFCCD instrument on the
Las Campanas 2.5~m du Pont telescope in 2019 to determine its
redshift.  By scaling the host spectrum and subtracting it from the SN~Ia observation
so as to make the obvious stellar \ion{Na}{1}~D blend at $\sim$5892~\AA\ absorption and the TiO feature
at $\sim$7150~\AA\ disappear, the other stellar features such as \ion{Ca}{2} H\&K, 
the G-band at $\sim$4300~\AA, H$\beta$, and \ion{Mg}{1}b $\lambda$5175 also mostly
disappeared (see Figure~\ref{fig:CSP13aao}).  We therefore have used this host-galaxy subtracted spectrum to derive
the spectroscopic properties at maximum light for this SN~Ia.

{\bf ASASSN-15eb}: The classification report by \citet{childress2015} does not refer to any peculiarities, however 
SNID yields some matches with 91T-like SNe at maximum light. According to our light
curve, this spectrum corresponds to phase $+4.5$ days. The pWs are indeed small, but this
is clearly caused by strong host contamination. 
Also strong Galactic \ion{Na}{1}~D absorption is evident in that spectrum.
The CSP 
spectrum published here, obtained at phase $+11.0$ days also exhibits significant host galaxy contamination
as well as strong \ion{Na}{1}~D absorption from the Milky Way. 
Both spectra show absorption minima on the red side of \ion{Si}{2} $\lambda$6355 which could be attributed
to \ion{C}{2}.

\section {Results}

In this section, we combine the \ion{Si}{2}
expansion velocities and selected pW measurements from this paper with those derived by \citet{folatelli2013}
to take a second look at some of the plots and correlations discussed in that paper.
In particular, our interest is to highlight agreements and differences between the properties
of SNe~Ia discovered in targeted versus untargeted searches.

\subsection {Temporal evolution of the expansion velocities of \ion{Si}{2} $\lambda$6355}
\begin{figure}
\plottwo{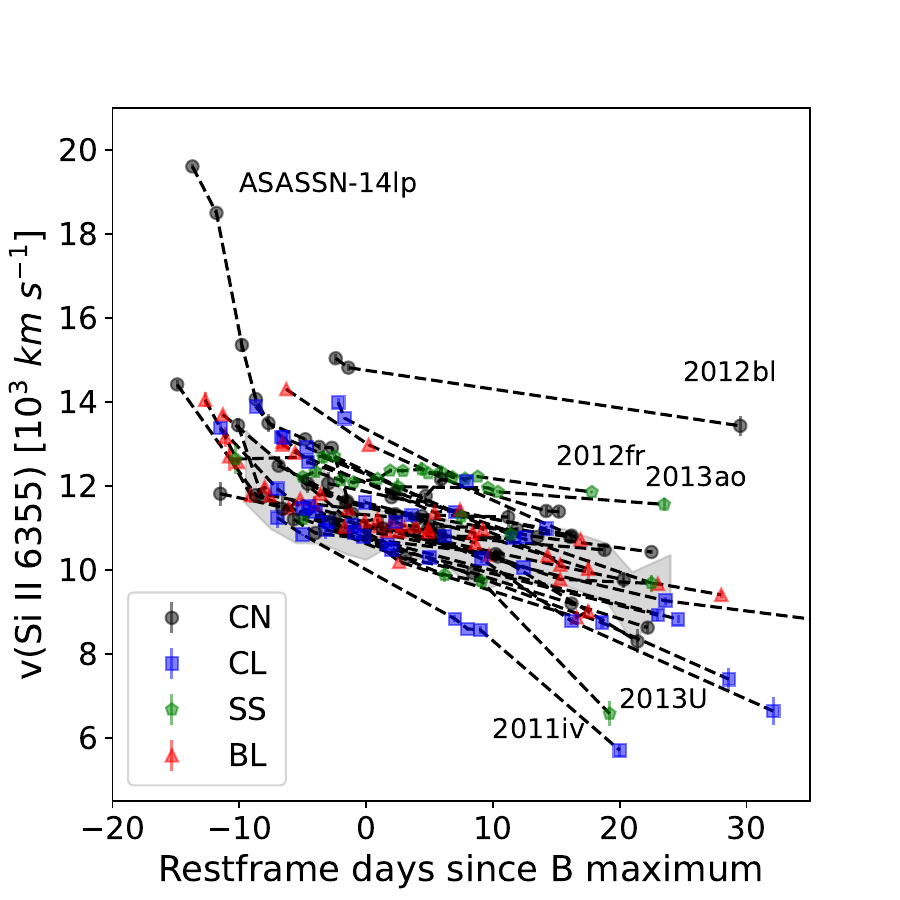}{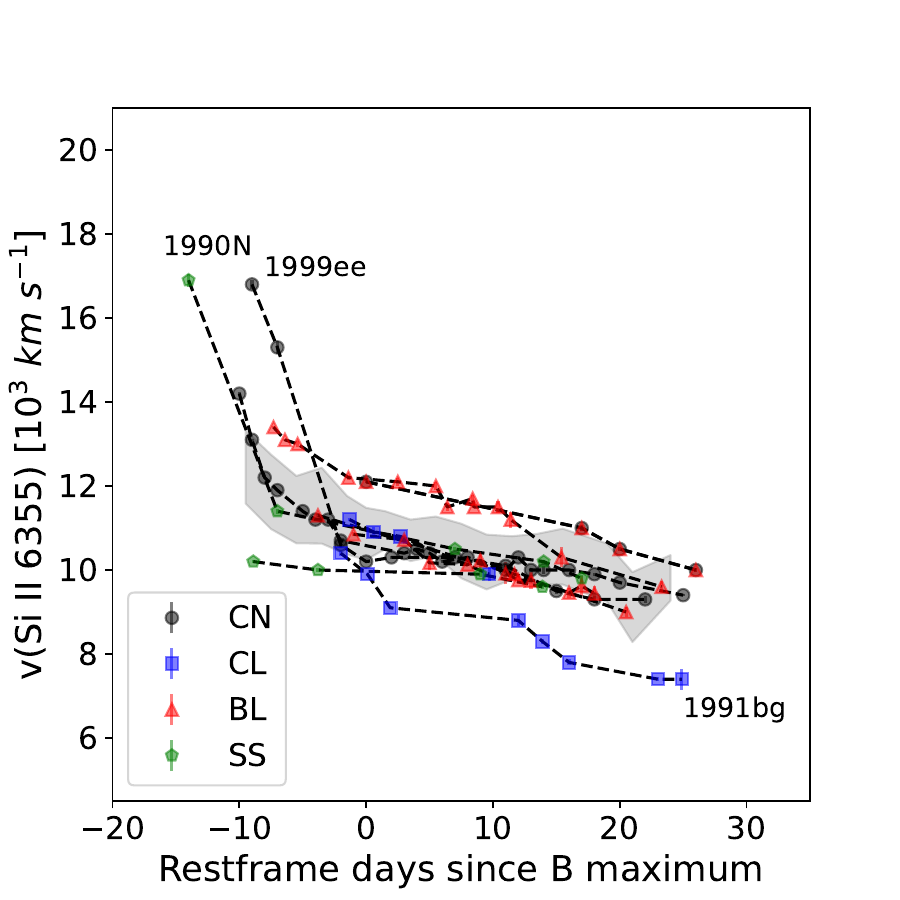}
\caption{Temporal evolution of the expansion velocities of the \ion{Si}{2} $\lambda$6355
line for samples of the CSP II targets (left) and the historical SNe~Ia (right).
The symbols reflect the corresponding Branch types: black circles for CN,
green pentagons for SS, red triangles for BL and blue squares for CL.
Error bars are plotted except when smaller than the symbols. Dashed lines connect data for each SN.
In both panels, the shaded region represents the upper and lower $1\sigma$ dispersion computed for all the Branch CN SNe with ``normal'' Wang
classification subtypes in the whole CSP I~\&~II sample.
\label {fig:SiII_evolution}}
\end{figure}

For a limited number of CSP-II targets and historical SNe Ia, the available data span enough time
to follow the evolution of the \ion{Si}{2} $\lambda$6355  expansion velocity 
to at least 20 days past maximum light. 
These observations are presented in the left and right panels of Figure~\ref{fig:SiII_evolution}, respectively, with the different Branch types  
indicated by the colors and shapes of the symbols.
In general, albeit with less data here, the behavior observed in this figure
is very similar to that of Figure 9 of \cite{folatelli2013}.
In both panels, the shaded region represents the upper and lower $1\sigma$ dispersion of the \ion{Si}{2} $\lambda$6355 expansion velocity evolution for the whole CSP I~\&~II sample of Branch CN SNe with ``normal'' Wang classification subtypes (see Table~7 for details). 

From Figure~\ref{fig:SiII_evolution} (left panel), we can see that SN\,2012bl shows high \ion{Si}{2} $\lambda$6355 velocity
which persists at 29 days past maximum light, although measurements of the minimum of \ion{Si}{2} $\lambda$6355 after
20 days post maximum are questionable due to blending with other features.  
SN\,2012fr \citep{childress2013,contreras18,cain2018} is an example of ``flat'' velocity evolution, in which the expansion velocity of 
\ion{Si}{2} $\lambda$6355 is almost constant over the period covered by our data ($-5.0$ to +17.8 days).

\begin{figure}
\plottwo{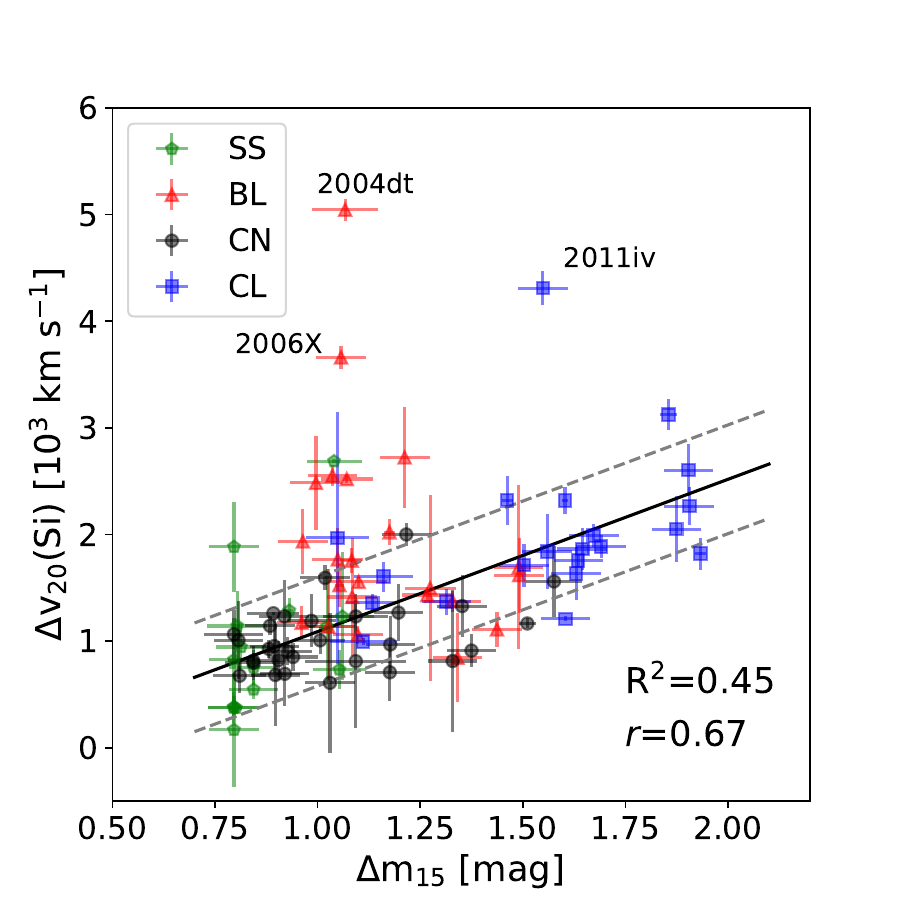}{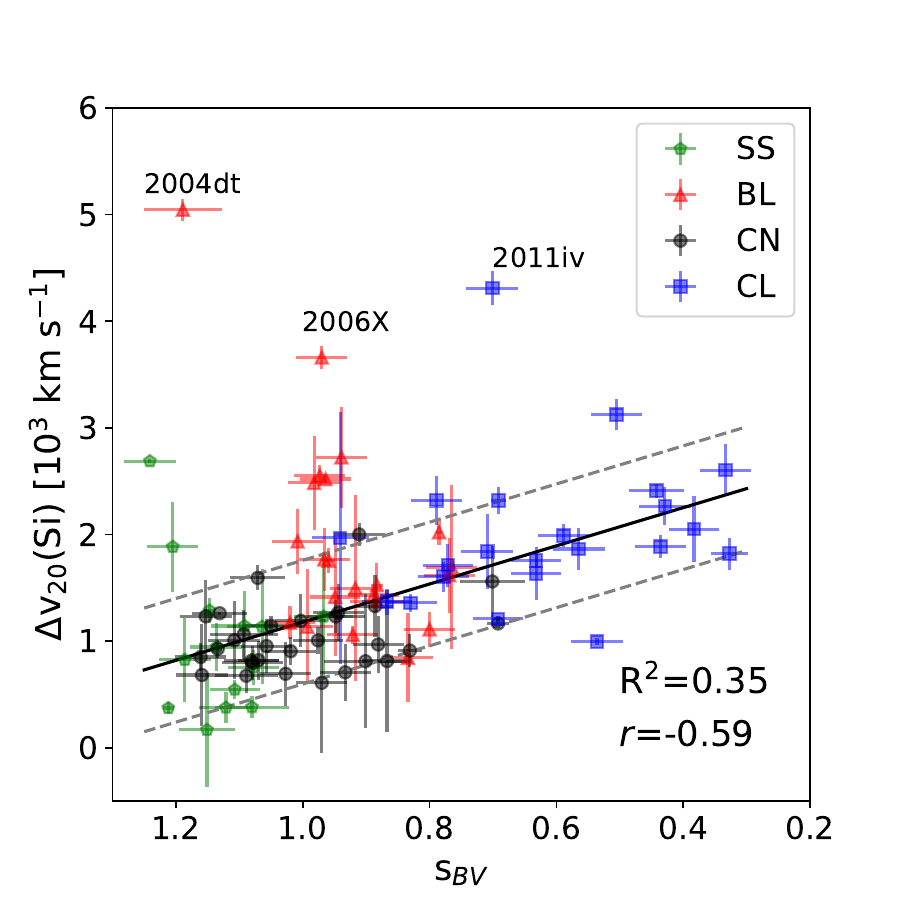}
\caption{ The left and right panels show the \ion{Si}{2} $\lambda$6355 velocity decline rates as a funcion of the light curve decline rate $\Delta{\rm m}_{15}$ 
and color stretch s$_{BV}$, respectively, for all the SNe Ia with sufficient phase coverage in the combined sample which includes CSP I, CSP II and
historical SNe Ia. The limited number of CSP II and historical targets for which this computation was possible causes the left panel of this
figure to be very similar to Fig. 21 of \citet{folatelli2013}.
As found in that work, the correlation is stronger when the BL SNe Ia are excluded
from the fit.
Shown in both figures are the best fit lines, intrinsic scatter lines (dotted grey),   
coefficients of determination (R$^2$), and correlation coefficients ($r$) excluding the BL events.
The transitional SN Ia 2011iv is an obvious outlier.}
\label {fig:SiII_decline_rate}
\end{figure}

For all the SNe~Ia presented in this paper with sufficient time coverage, the difference between the \ion{Si}{2} $\lambda$6355 velocity at maximum light and at
20 days past maximum, $\Delta v_{20}$(\ion{Si}{2}), was calculated using the same methodology described in \S3.1.1 of \citet{folatelli2013}. These values are given in the
last column of Table~\ref{vels} and plotted in Figure~\ref{fig:SiII_decline_rate} as a function of light-curve decline rate $\Delta{\rm m}_{15}$ and color-stretch s$_{BV}$, respectively, along with the CSP-I objects already presented in \citet{folatelli2013}.  Figure~\ref{fig:SiII_decline_rate} confirms the strong correlation \footnote{All the correlations presented in this work have been computed via the {\it linmix} code 
(https://github.com/jmeyers314/linmix) based on \citet{kelly2007}. In the corresponding figures we present the data along with best fit lines, intrinsic scatter lines, coefficients of determination (R$^2$) and Pearson correlation coefficients ($r$).}
observed by  \citet{folatelli2013} between $\Delta v_{20}$(\ion{Si}{2}) and 
$\Delta{\rm m}_{15}$ for Branch SS, CN, and CL SNe, suggesting
that these events form a single sequence.
On the other hand, the lack of a correlation for the Branch BL
events is consistent with previous hints that these objects
may represent a distinct group of SNe~Ia \citep[e.g., see][]{wang2013,burrow2020}.

\subsection {Correlations involving spectroscopic parameters}

In the analysis of spectroscopic parameters at maximum light, we 
consider separately the objects discovered by targeted and untargeted
surveys.  That is, SNe~Ia in the CSP-I, CSP-II, and historical samples 
discovered by amateur astronomers or other targeted surveys (e.g., the Lick
Observatory Supernova Survey, LOSS; or the Chilean Automatic Supernova Search, CHASE)
will be considered as one sample, while all the targets 
drawn from untargeted surveys such
as the La Silla Quest (LSQ), the Palomar Transient Factory (PTF, iPTF),
the Sloan Digital Sky Survey (SDSS) or the Catalina Real Time Transient Survey (CRTS), ASAS-SN, among others, will be considered as a second sample.
The Cal\'{a}n/Tololo SNe are a special case since the photographic plates were taken in an untargeted fashion, but were searched for stellar objects that appeared near galaxies.
We therefore have grouped these SNe with the targeted events.
Most of the CSP-I SNe~Ia belong to the {\bf targeted} group and most of the CSP-II SNe~Ia
belong to the untargeted one. However, there are a few exceptions.
CSP-II SNe~Ia for which we derived spectroscopic parameters
at maximum light, and that are included in the first group
of targeted survey discoveries are: PSN J13471211-2422171 
and the SNe 2011iv, 2011jh, 2012E, 2012ah, 2012fr, 2012hd, 
2012hr, 2012ht, 2013aa, 2013fz, 2013gy, 2013hn, 2014I,
2014Z, 2014ao, 2014at, 2014dn, 2014eg and 2015F. On the other hand, CSP-I
targets for which we present spectroscopic parameters at maximum light that were discovered by  untargeted searches are SNe~2007if, 2007ol, 2008bz and 2008fr.

\subsection {The Branch Diagram}

\begin{figure}
\plottwo{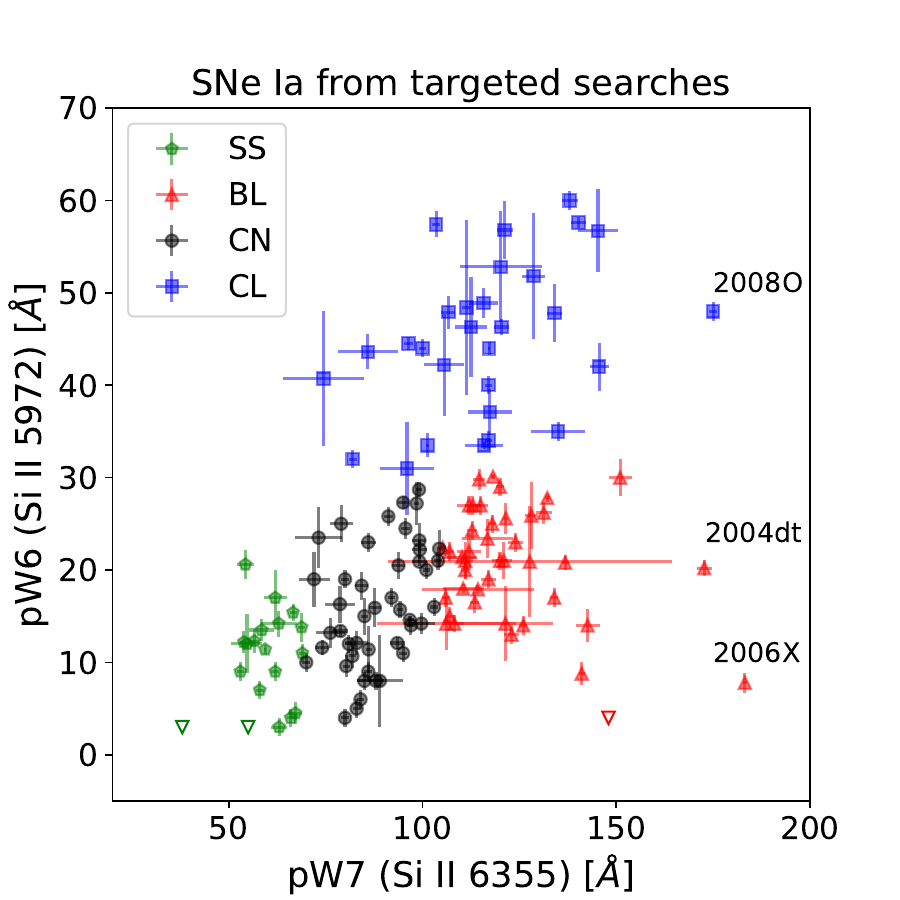}{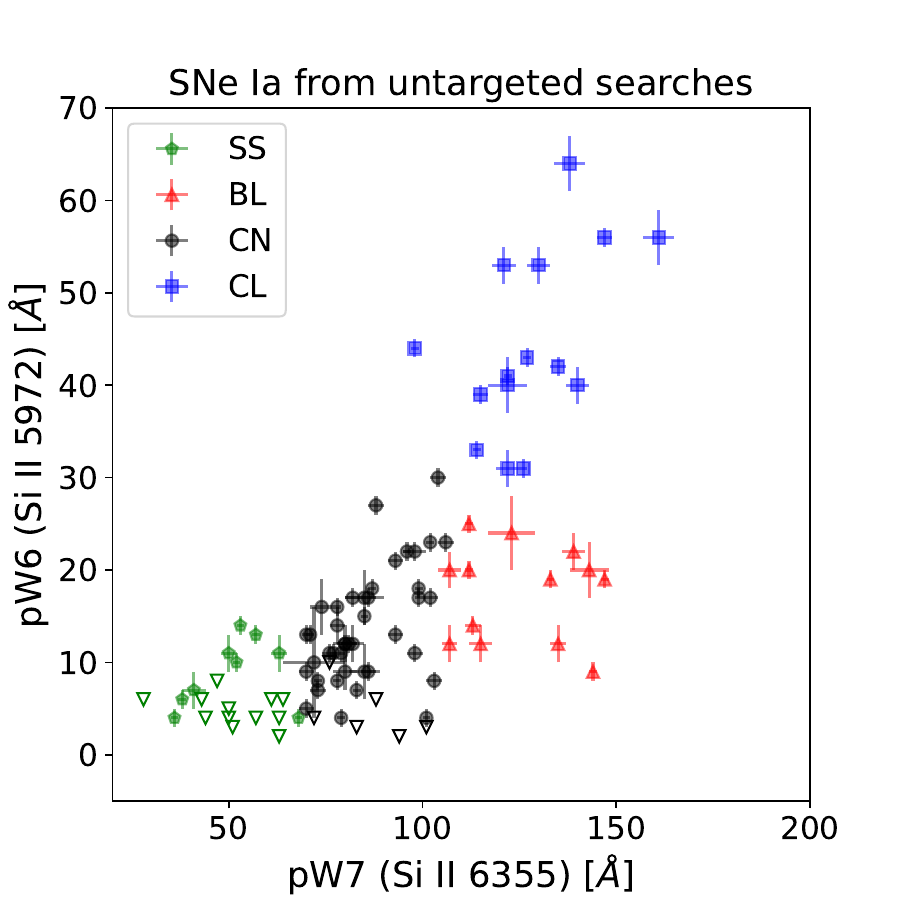}
\caption{Branch diagrams for our two different samples. Left: Historical SNe~Ia and CSP I+II SNe~Ia discovered during targeted surveys
(45 CN, 29 CL, 42 BL and 20 SS SNe Ia). Right: CSP I+II SNe Ia from untargeted surveys (48 CN, 15 CL, 13 BL and 21 SS SNe Ia).
The meaning of the symbols is as follows: black dots represent CN SNe~Ia; green pentagons represent SS SNe~Ia; 
red triangles represent BL SNe~Ia and blue squares represent CL SNe~Ia. 
Open symbols represent upper limits.
\label {fig:Branch}
}
\end{figure}

\begin{figure}
    \plottwo{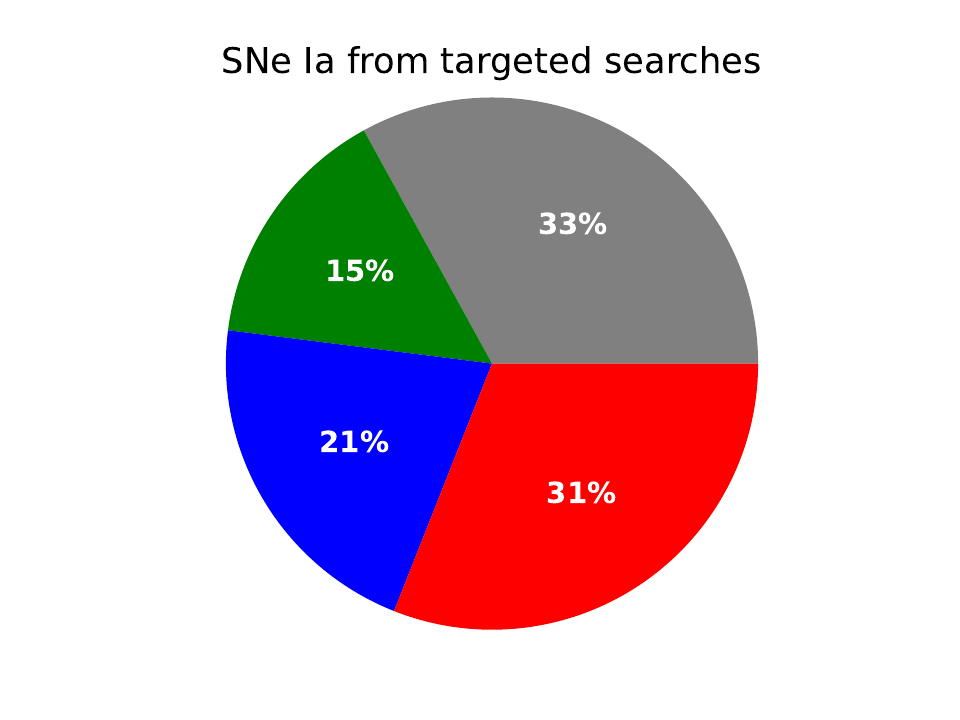}{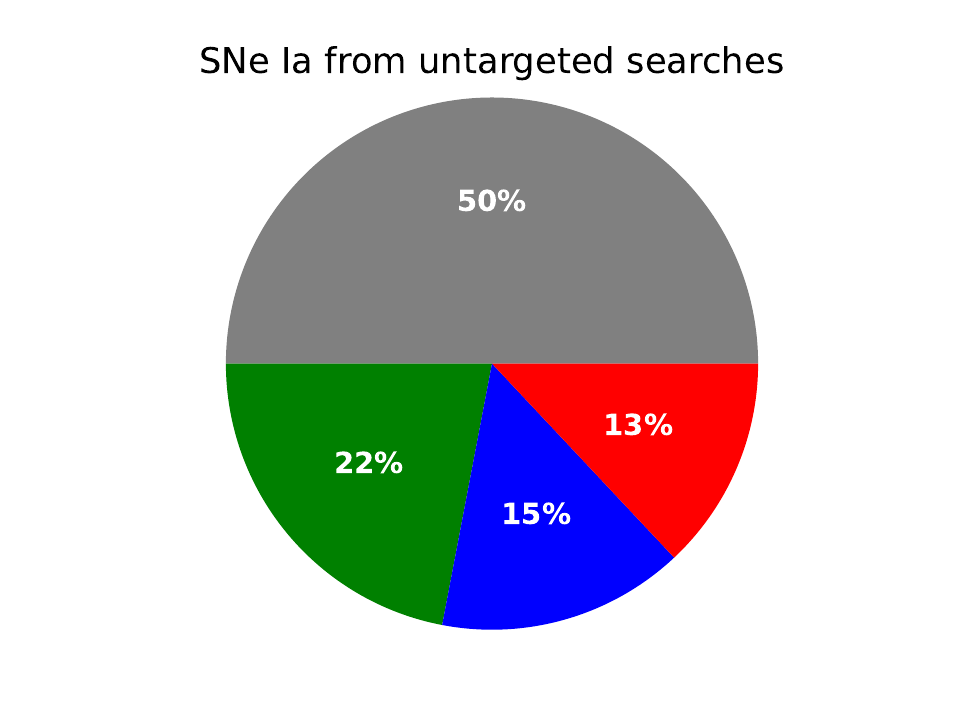}
    \caption{Pie charts showing the incidence of the different Branch types of SNe Ia (CN in grey, SS in green, CL in blue and BL in red) in our targeted and untargeted samples, respectively.
    \label{fig:pie_chart}}
\end{figure}

\begin{figure}
    \centering
    \includegraphics[width=0.8\textwidth]{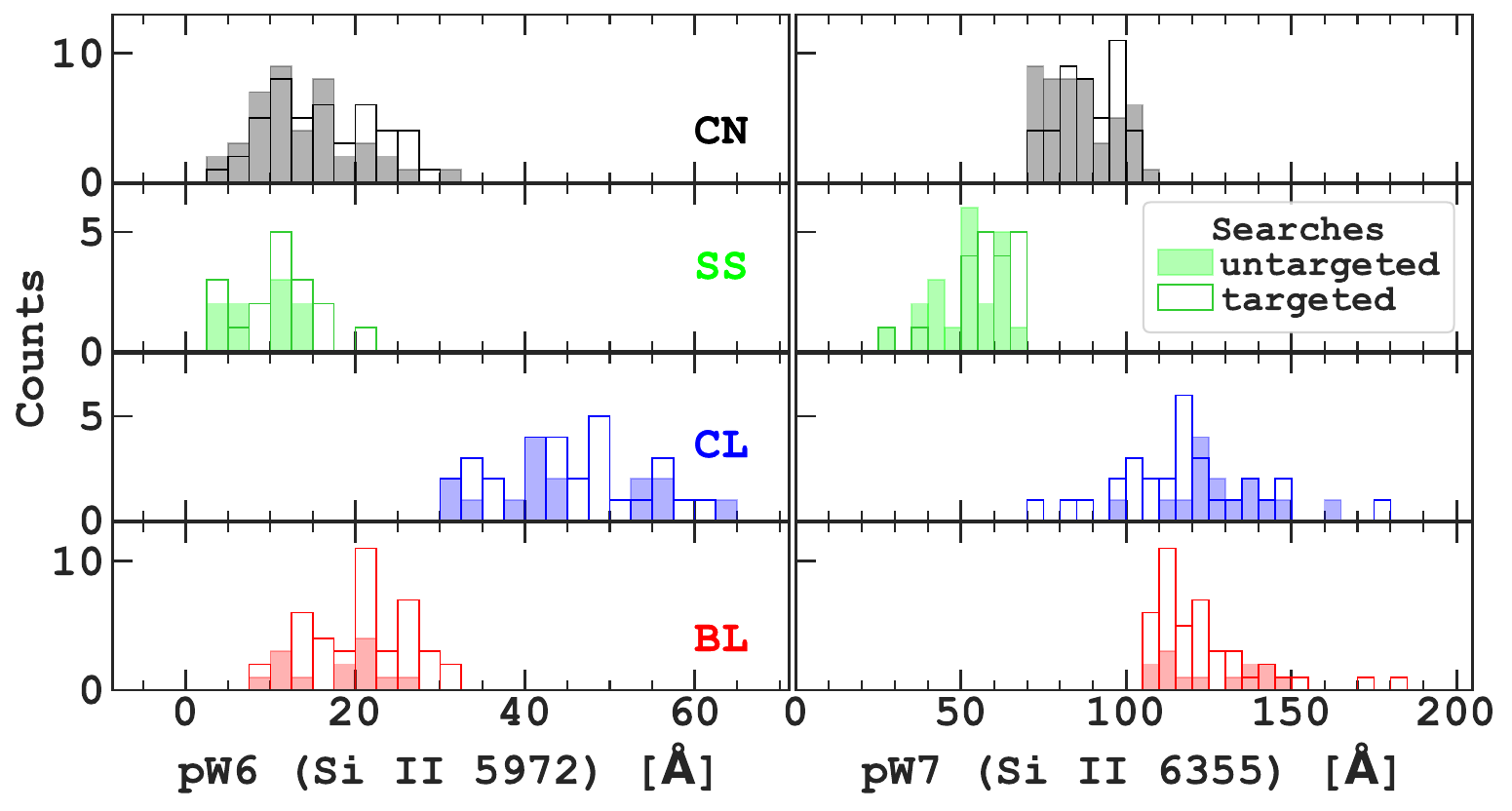}
    \caption{Histograms showing the distribution of the measured pW6 and pW7 separately for each Branch type. Colored and empty bars 
    represent untargeted and targeted searches, respectively.
    \label{fig:pW6_pW7_histogram}}
\end{figure}

Figure~\ref{fig:Branch} displays the \citet{branch2006} diagram for the targeted (left) and
untargeted (right) samples of CSP and historical SNe Ia.  In defining the boundaries between the four
classes (CN = ``core normal'', SS = ``shallow silicon'', BL = ``broad lined'', CL = ``cool''), we follow
the definitions adopted by \citet{folatelli2013}.
The largest difference between these diagrams 
is for the BL SNe~Ia, whose relative numbers are clearly different.  To be precise, the BL SNe~Ia represent 31\% $\pm$ 5\% of the targeted sample, 
but only 13\% $\pm$ 4\% of the untargeted events.  This difference is explained by the fact that targeted searches are biased
towards luminous galaxies, and high-velocity SNe Ia (approximately two-thirds of which belong to the
BL class) are known to occur preferentially in luminous galaxies \citep{wang2013}. 
Similarly to the BL events, CL SNe --those with relatively large values of pW6 (\ion{Si}{2} 5972)-- represent 21\% $\pm$ 4\% of the targeted, and 15\% $\pm$ 4\% of the untargeted samples, respectively. 
As for the CN SNe, they are more frequent in the untargeted sample, amounting to 50\% $\pm$ 9\% 
of the total, compared to 33\% $\pm$ 6\% of the targeted sample.
The SS SNe, which correspond to the classical 1991T-like and 1999aa-like events, show a less significant difference, amounting to 22\% $\pm$ 5\% of the untargeted sample, while in the targeted sample they represent 15\% $\pm$ 4\%. These percentages are illustrated for the targeted and  untargeted samples, respectively, in Figure~\ref{fig:pie_chart}.
For further comparison, we show in Figure~\ref{fig:pW6_pW7_histogram}  histograms of our  pW6 and pW7 measurements at maximum light for the 
untargeted and targeted samples, separately for each of the Branch types.

Another way to look at these numbers is to consider ratios. In particular, how do the ratios of the numbers of SS, CL and BL SNe~Ia compare to the number
of CN SNe~Ia for the targeted and untargeted samples? In the case of the targeted SNe, N(SS)/N(CN)=0.44, N(CL)/N(CN)=0.64, 
and N(BL)/N(CN)=0.93, while for the untargeted events we find N(SS)/N(CN)=0.44, N(CL)/N(CN)=0.31, and N(BL)/N(CN)=0.27. 
These numbers imply that SS SNe~Ia are equally common with respect to {\bf CN} SNe in targeted vs. 
 untargeted surveys. On the other hand, CL SNe~Ia are twice as common and BL SNe are more than three times more common with respect to {\bf CN} SNe in targeted surveys compared to untargeted
surveys. This likely reflects the fact that SS and CN SNe~Ia occur in hosts over a large range of luminosity, whereas CL and BL SNe~Ia prefer luminous hosts \citep[e.g., see][]{neill09,wang2013,pan2020}.

Note that \citet{burrow2020} explored the Branch diagram through a cluster analysis instead of using pre-defined group boundaries as we have done in Figure~\ref{fig:Branch}.
Comparing the Branch type classifications in Tables~\ref{CSP-prop} and \ref{CT-prop} for the 43 SNe~Ia in common with the sample that \citeauthor{burrow2020} analyzed using a two-dimensional Gaussian mixture model (2D~GMM),
60\% are in agreement.  Not surprisingly, the objects for which the classifications differ lie at or near the borders of the Branch groups.

\begin{figure}
    \begin{center}
    \includegraphics[width=0.6\textwidth]{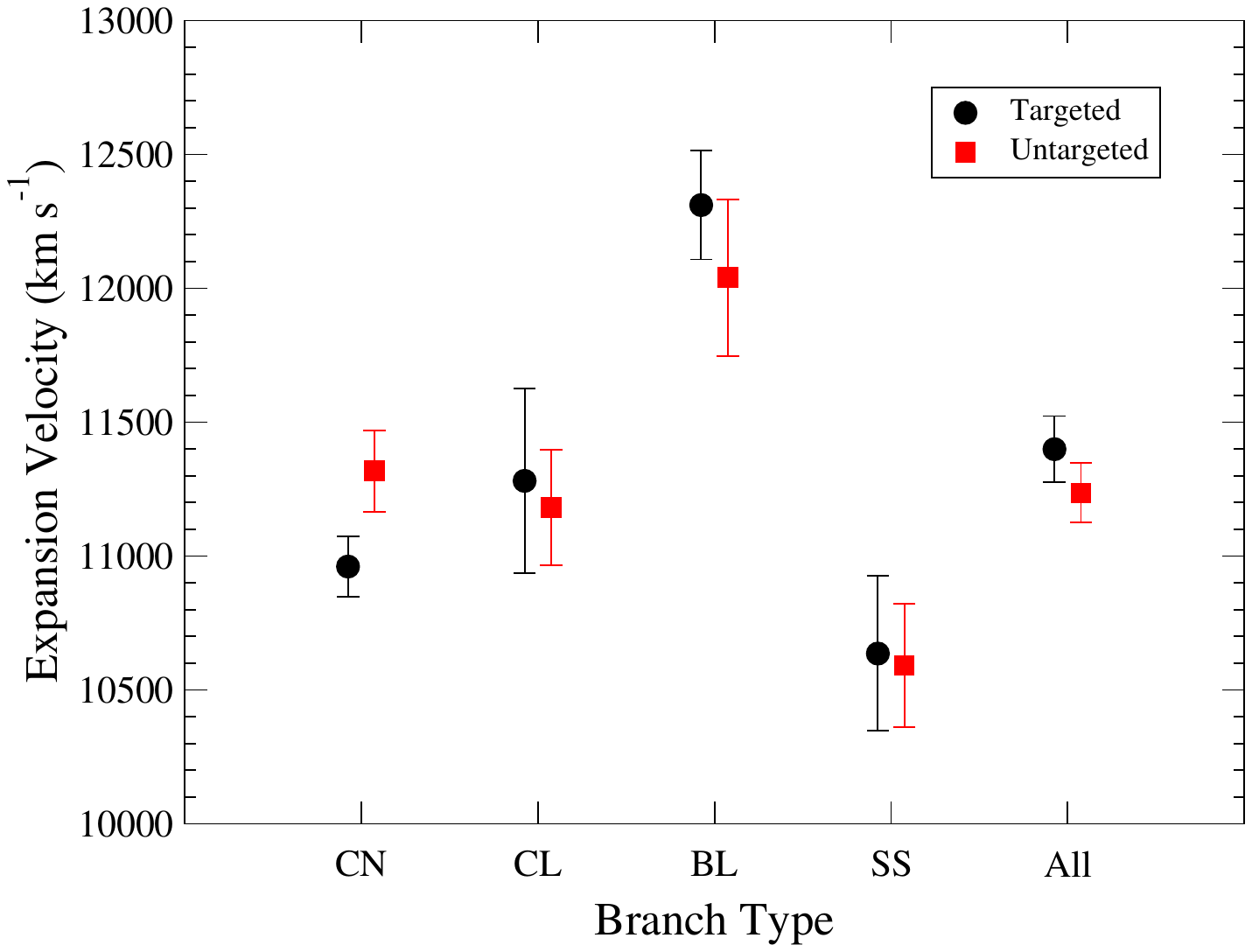}
    \end{center}
    \caption{Average expansion velocities of the \ion{Si}{2} $\lambda$6355 for each of the Branch types in the targeted and untargeted
    samples, and for each of the two samples together. The error bars represent the standard errors of the mean.
    \label{fig:vels_vs_Branch}}
\end{figure}

Some studies have found correlations between SNe Ia ejecta velocities and their host properties  (e.g. \citealt{pan2015, pan2020, dettman2021}).
Considering that their results imply that high-velocity SNe Ia tend to prefer high-stellar-mass hosts, more frequent in targeted searches, 
we compared the \ion{Si}{2} $\lambda$6355 expansion velocities at maximum light derived for our targeted and untargeted samples as a whole, 
as well as for the different Branch types within each sample.
While the expansion velocities tend to be higher for BL SNe Ia, and somewhat lower for SS SNe Ia, the differences between the targeted and untargeted sample
are insignificant considering the uncertainties. Figure~\ref{fig:vels_vs_Branch} presents the average velocities
at maximum light for \ion{Si}{2} $\lambda$6355 as a function of the different Branch types, and for the whole targeted and untargeted samples,
respectively.

\subsection {Correlations between pW values of different features at maximum light}

\begin{figure}
\plottwo{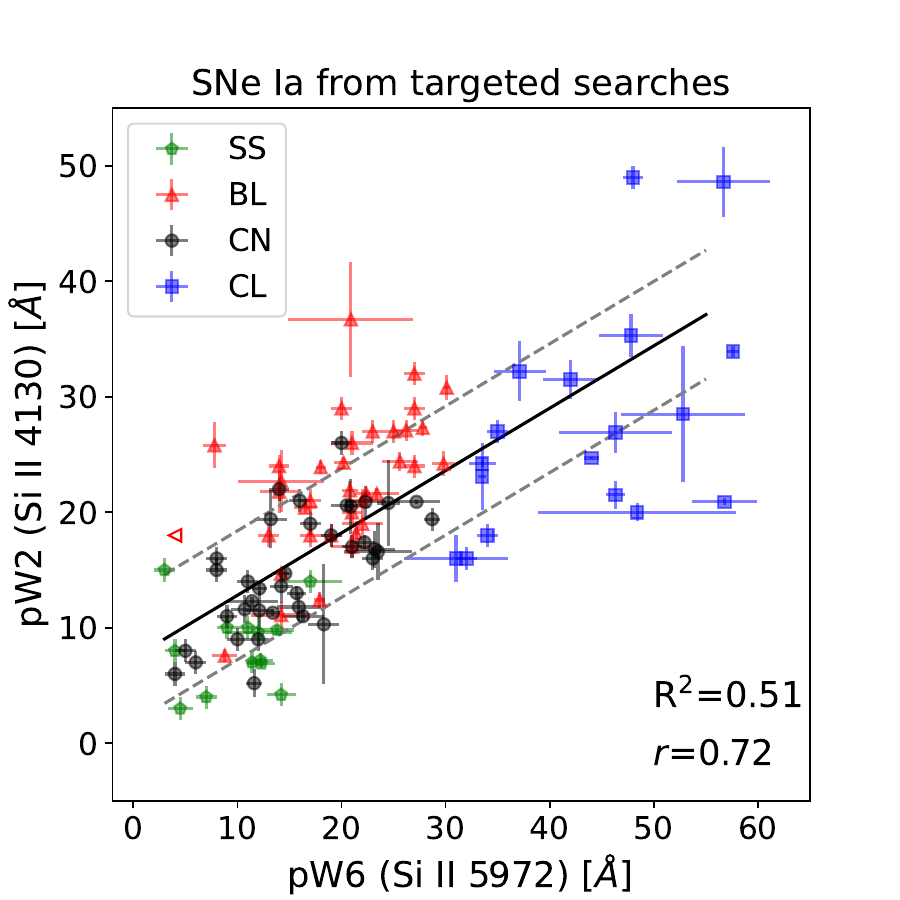}{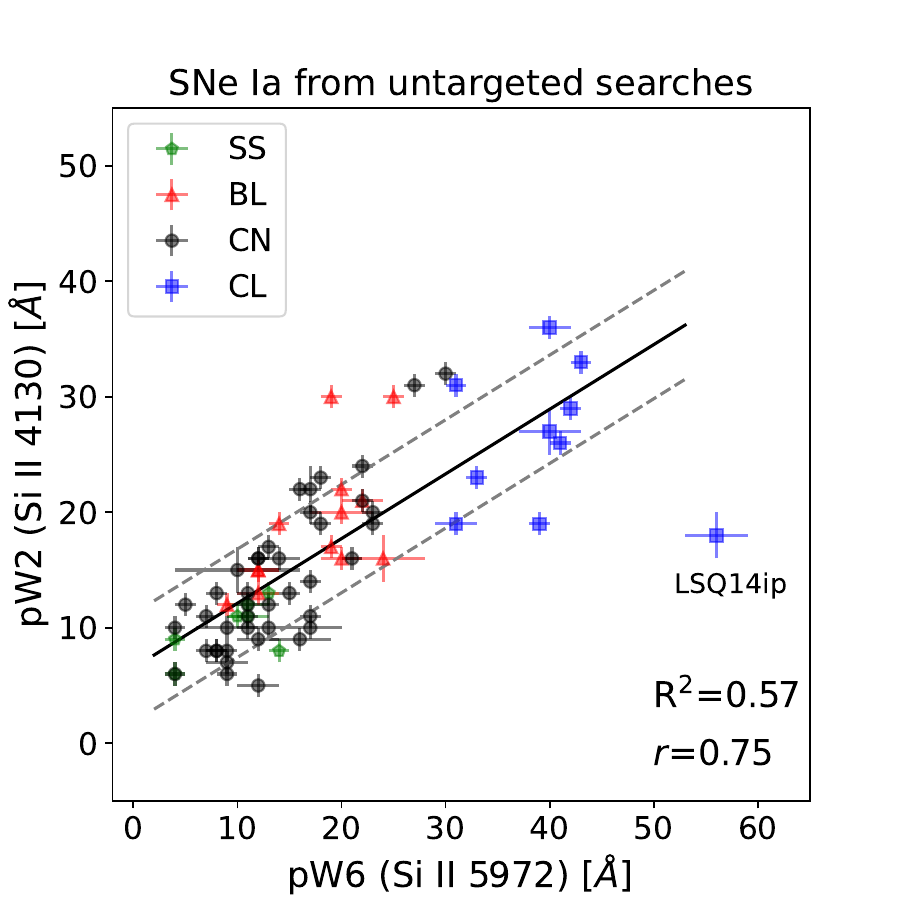}
\caption{Correlation between pW values of \ion{Si}{2}~$\lambda$4130 and \ion{Si}{2}~$\lambda$5972 for our two different samples:
historical SNe Ia and CSP SNe Ia from targeted searches (left) and CSP SNe Ia discovered by untargeted searches (right). 
The meaning of the symbols is as in Figure~\ref{fig:Branch}. The correlation strengths are similar for both samples as well as the slopes of the best-fit lines {\bf (0.54$\pm$0.05 and 0.56$\pm$0.06} for the targeted and untargeted samples, respectively).
\label{fig:pW2_vs_pW6}}
\end{figure}

\begin{figure}
\plottwo{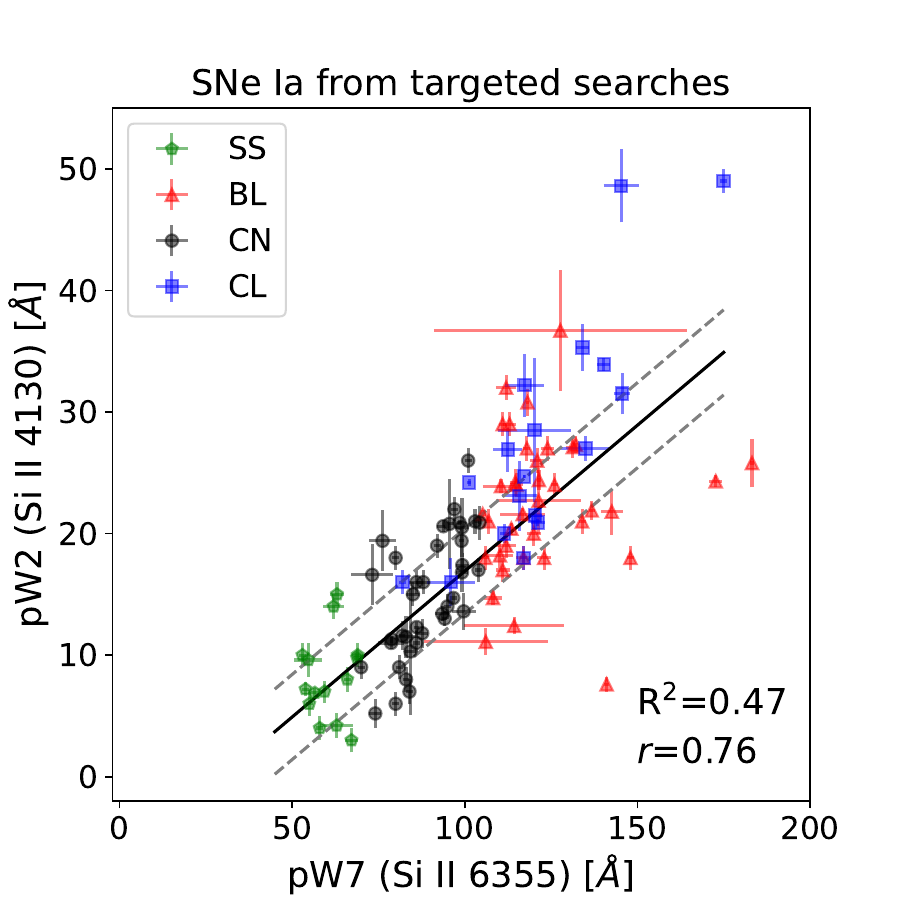}{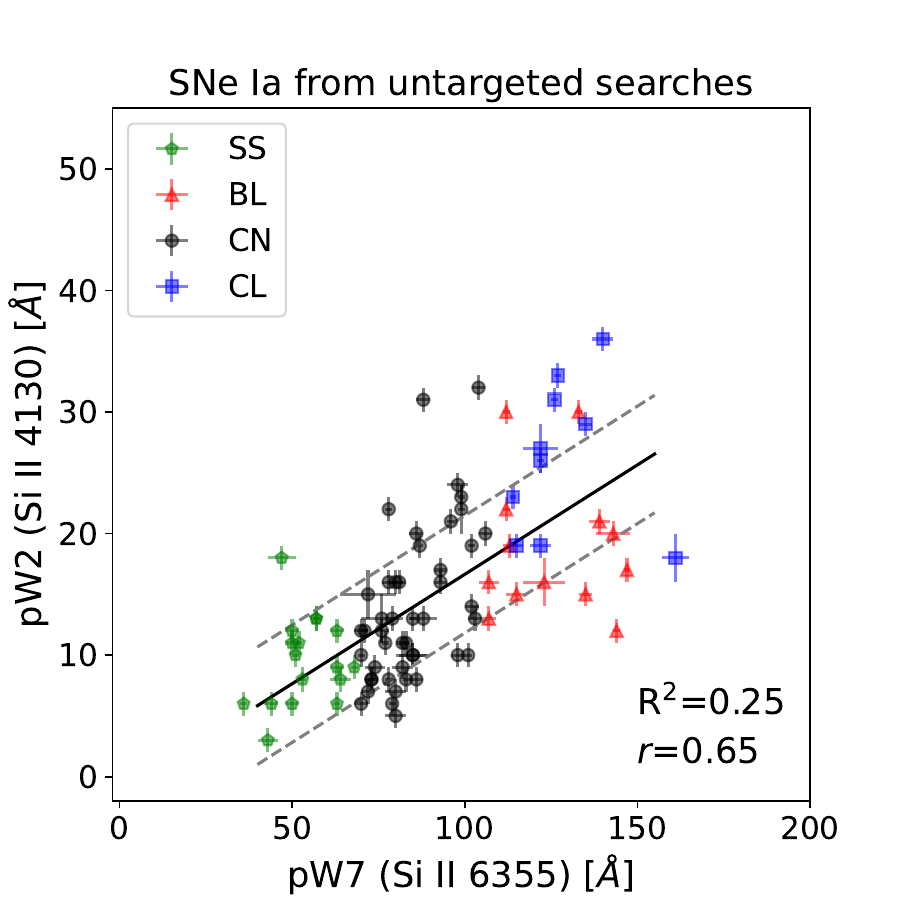}
\caption{Correlation between pW values of \ion{Si}{2}~$\lambda$4130 and \ion{Si}{2}~$\lambda$6355 
for our two different samples of SNe~Ia discovered by targeted (left) and untargeted (right) searches.
The meaning of the symbols is as in Figure~\ref{fig:Branch}. 
The correlation is stronger and has a smaller intrinsic scatter for the targeted sample.
The best-fit lines have slopes of 0.24$\pm$0.04 and 0.18$\pm$0.04 for the targeted and untargeted samples, respectively.}
\label{fig:pW2_vs_pW7}
\end{figure}

\begin{figure}
\plottwo{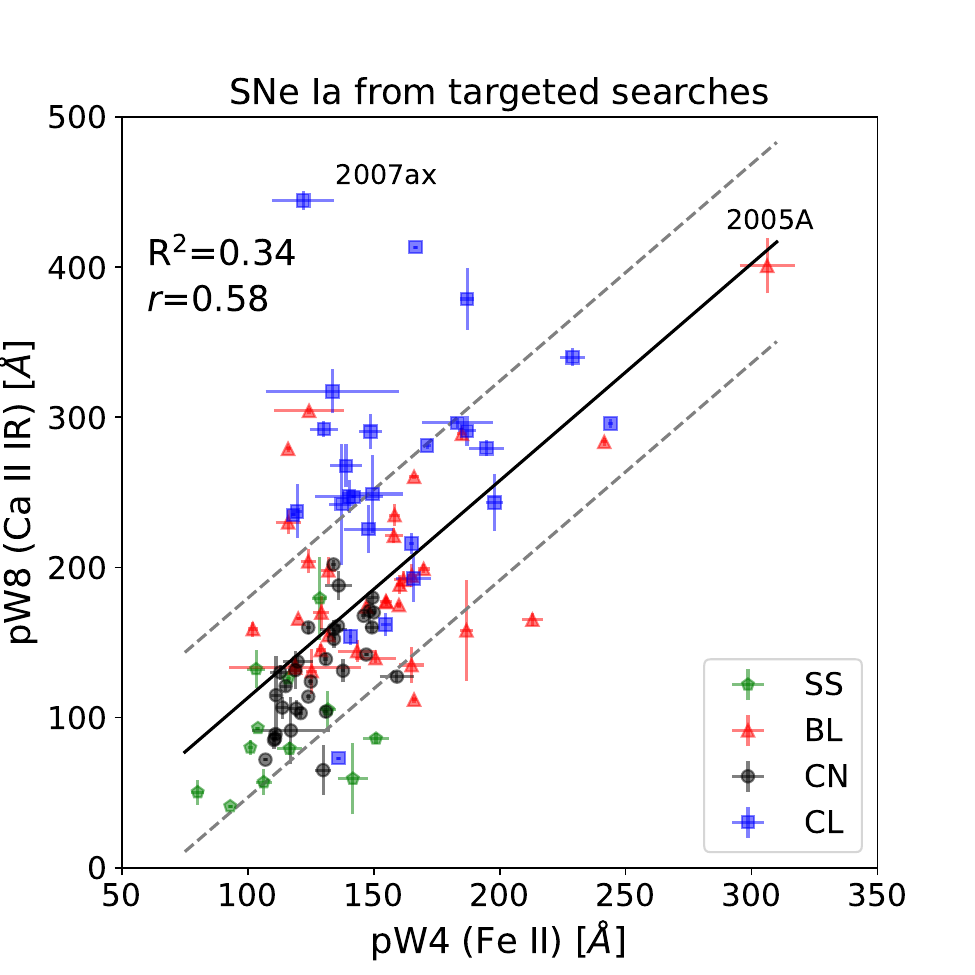}{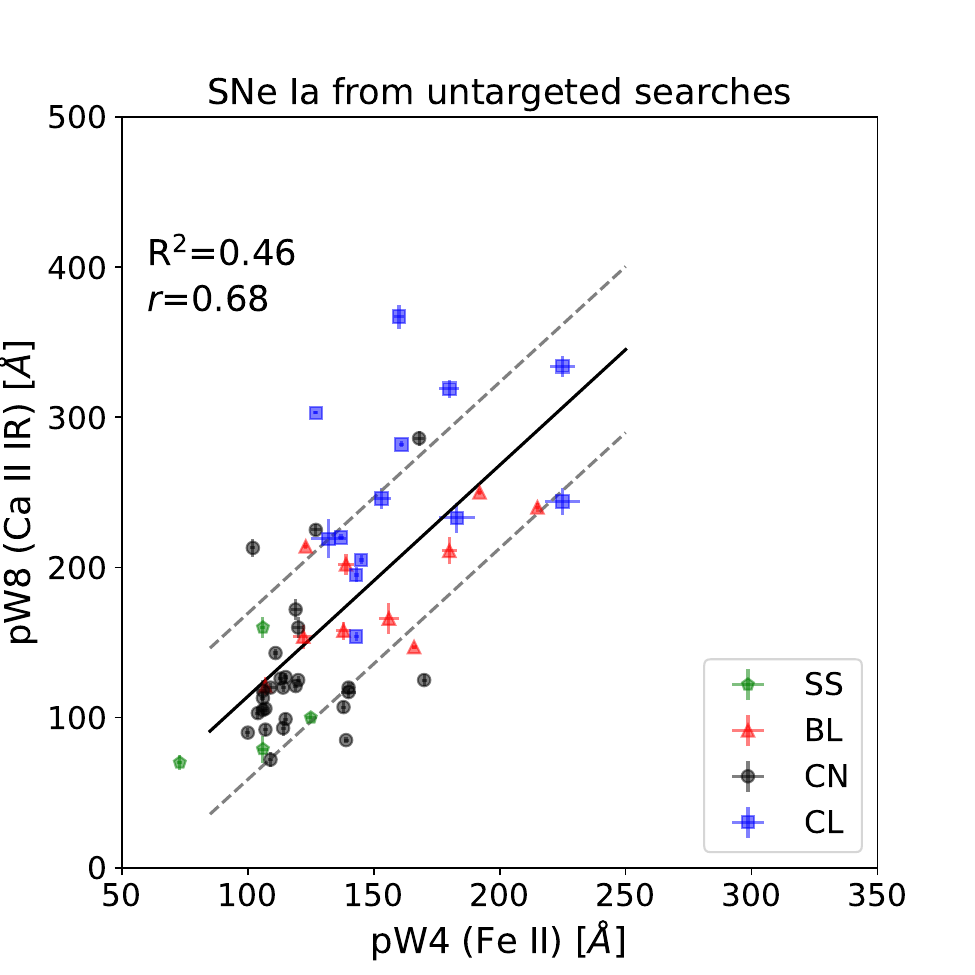}
\caption{Correlation between pW values of the \ion{Ca}{2}~IR triplet and the
\ion{Fe}{2} feature at $\sim$4600 \AA\, 
for our two different samples of SNe~Ia discovered by targeted (left) and untargeted
(right) searches.
The meaning of the symbols is as in Figure~\ref{fig:Branch}. The correlation is a bit stronger for the untargeted sample, while the slopes
are indistinguishable (1.51$\pm$0.21 and 1.54$\pm$0.24 for the targeted and untargeted samples, respectively).}
\label{fig:pW8_vs_pW4}
\end{figure}

\begin{figure}
\plottwo{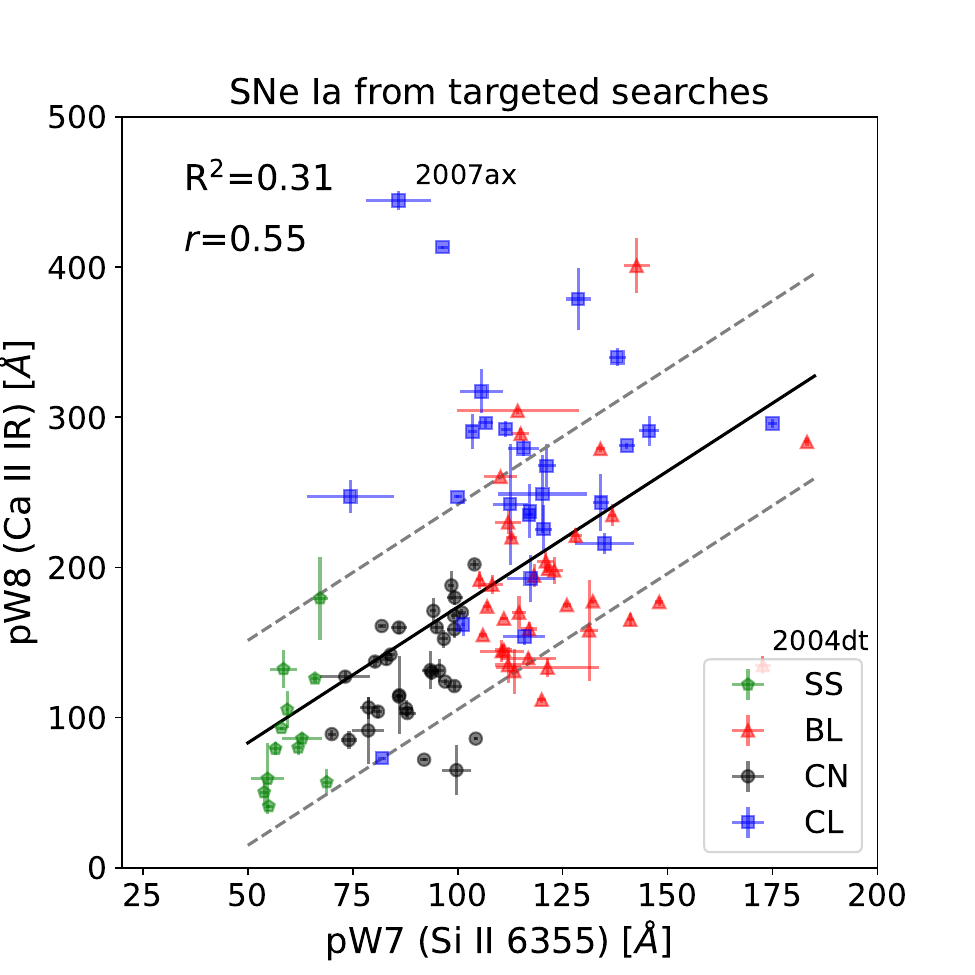}{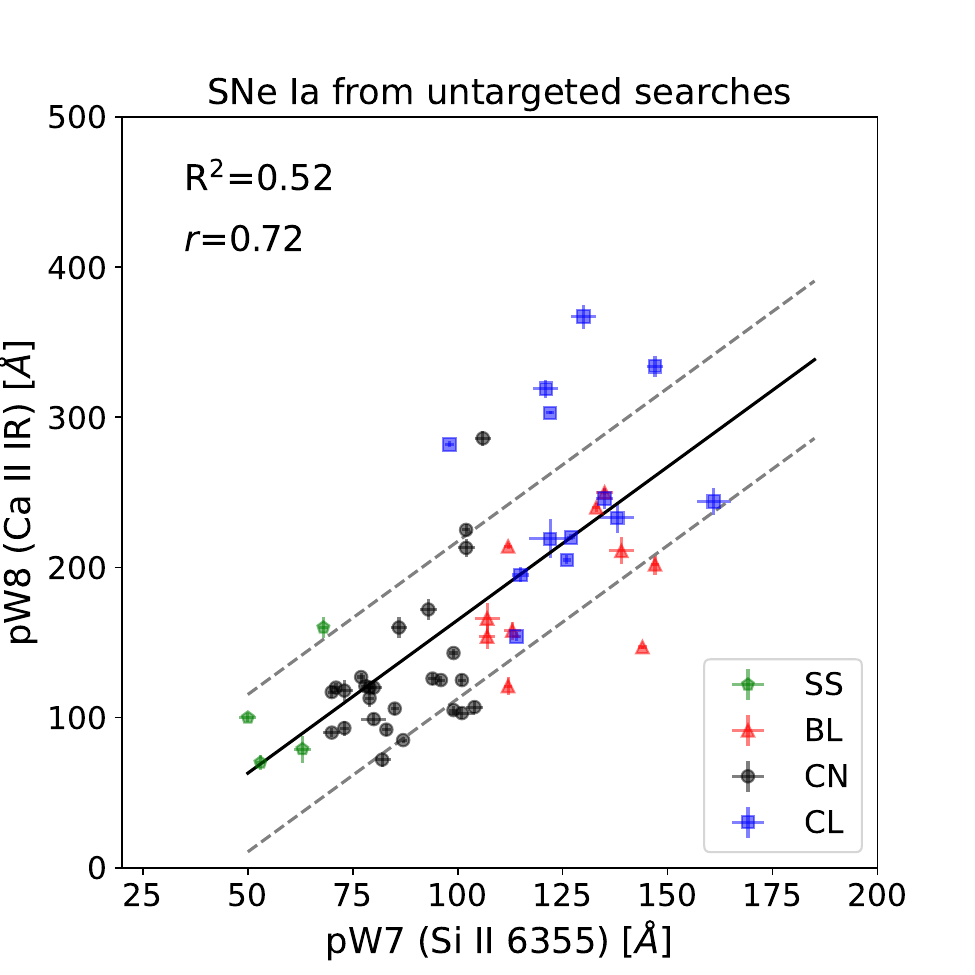}
\caption{Correlation between pW values at maximum light of the \ion{Ca}{2}~IR triplet  and \ion{Si}{2}~$\lambda$6355 for our two different samples of SNe~Ia discovered by targeted (left) and untargeted (right) searches.
The meaning of the symbols is as in Figure~\ref{fig:Branch}.
The slopes of the best-fit lines are equivalent within the uncertainties: 1.81$\pm$0.27 and 2.04$\pm$0.28 for the targeted and untargeted samples, respectively.
\label{fig:pW8_vs_pW7}}
\end{figure}

In their study of SNe~Ia from the CSP-I, \citet{folatelli2013} searched for correlations
between the different pW measurements at maximum light.  In Figures~\ref{fig:pW2_vs_pW6} to \ref{fig:pW8_vs_pW7},
we reproduce the four strongest correlations that they found, plotting each separately for the
targeted and untargeted samples.  \citeauthor{folatelli2013} found that the correlations were 
tighter (Pearson correlation coefficients $\rho > |0.75|$) if the high velocity and fast-declining
($v$(\ion{Si}{2}(6355) $<12,000$ km~s$^{-1}$ and $\Delta{\rm m}_{15}(B) < 1.7$ mag)
events were excluded from the fits, but in this paper, we do not make this distinction.
In each of the Figures~\ref{fig:pW2_vs_pW6} to \ref{fig:pW8_vs_pW7}, coefficients of determination (R$^2$) and
Pearson correlation coefficients (r) are shown near the top left or bottom right
corners of each panel.

As shown in the right panel of Figure~\ref{fig:pW2_vs_pW6}, the strongest correlation for the 
untargeted sample is found between the pW values for \ion{Si}{2}~$\lambda$4130 versus 
\ion{Si}{2}~$\lambda$5972.  In this plot, the CL SN~Ia LSQ14ip appears as an outlier in an otherwise
strong positive correlation.  However, the left panel of Figure~\ref{fig:pW2_vs_pW6} corresponding to the
targeted sample shows a large dispersion in the measurements for the CL SNe~Ia, with LSQ14ip lying within this dispersion.
LSQ14ip is an extremely cool SN~Ia, very similar to SN~1986G \citep{phillips1987}, which showed
strong \ion{Ti}{2} absorption at maximum light.  The \ion{Si}{2}~$\lambda$4130 line is blended with
the \ion{Ti}{2} absorption \citep[e.g., see][]{ashall2016}, making it difficult to measure an accurate
pW value. This blending undoubtedly accounts for the large dispersion in the CL events observed in the targeted sample. 

In Figure~\ref{fig:pW2_vs_pW7} we plot the pW values for the \ion{Si}{2}~$\lambda$4130
and \ion{Si}{2}~$\lambda$6355 absorptions.  In this case, the correlation is tighter
for the targeted sample.  As might be expected, however, correlations between all three of the
\ion{Si}{2} lines are similarly strong for both the targeted and untargeted samples.

Figure~\ref{fig:pW8_vs_pW4} displays the pW values for the \ion{Ca}{2}~IR triplet plotted versus
the pW4 (\ion{Fe}{2}) measurements.  
The correlation is tighter for the SNe Ia discovered in untargeted searches.
In Figure~\ref{fig:pW8_vs_pW7}  the correlation between the pW8 (\ion{Ca}{2} IR) and pW7 (\ion{Si}{2} 6355) parameters is displayed. 
We see that the source of much of the dispersion, notably in the plot corresponding to the targeted searches, comes from the CL SNe~Ia. 

\begin{figure}
 \plottwo{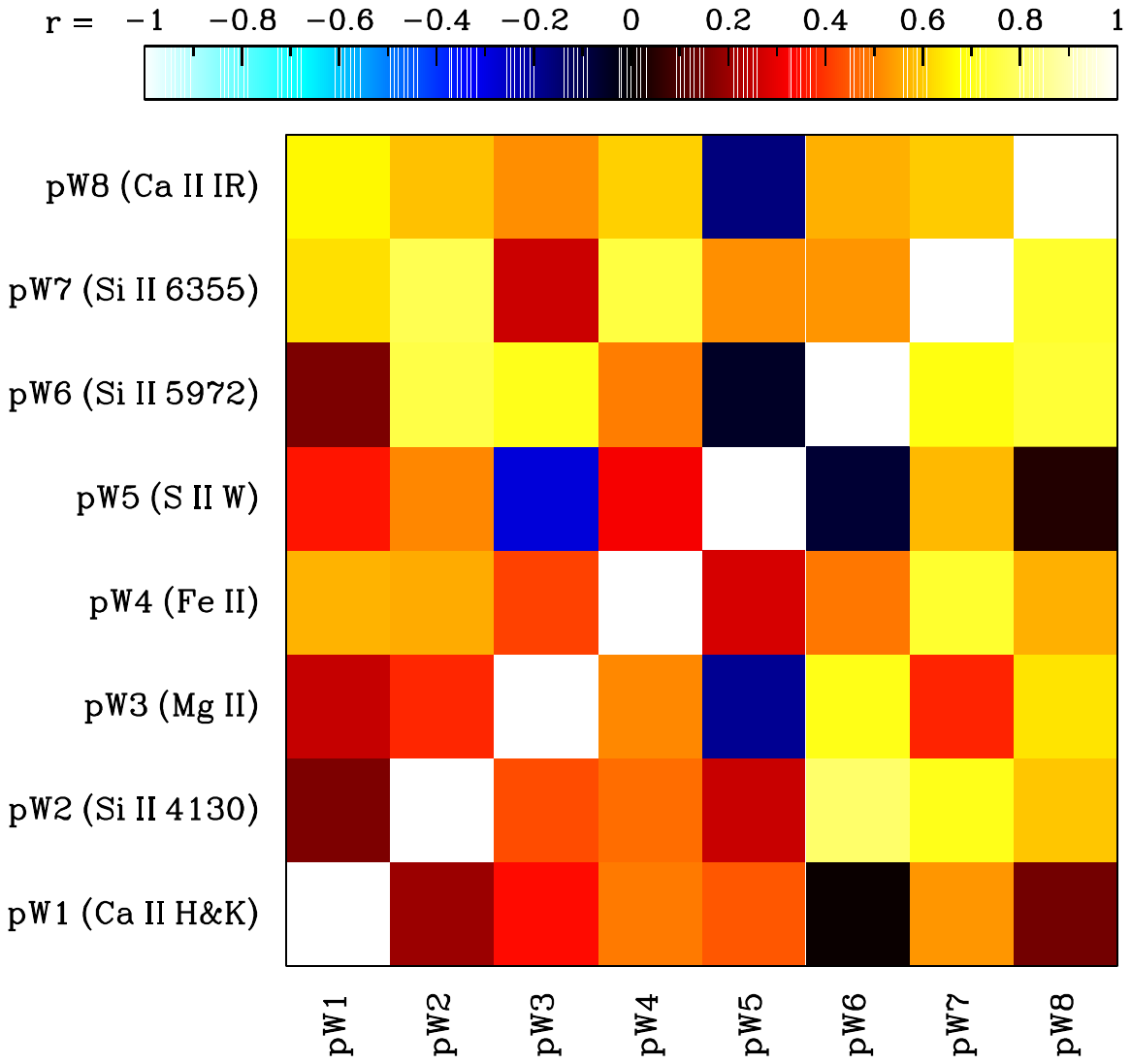}{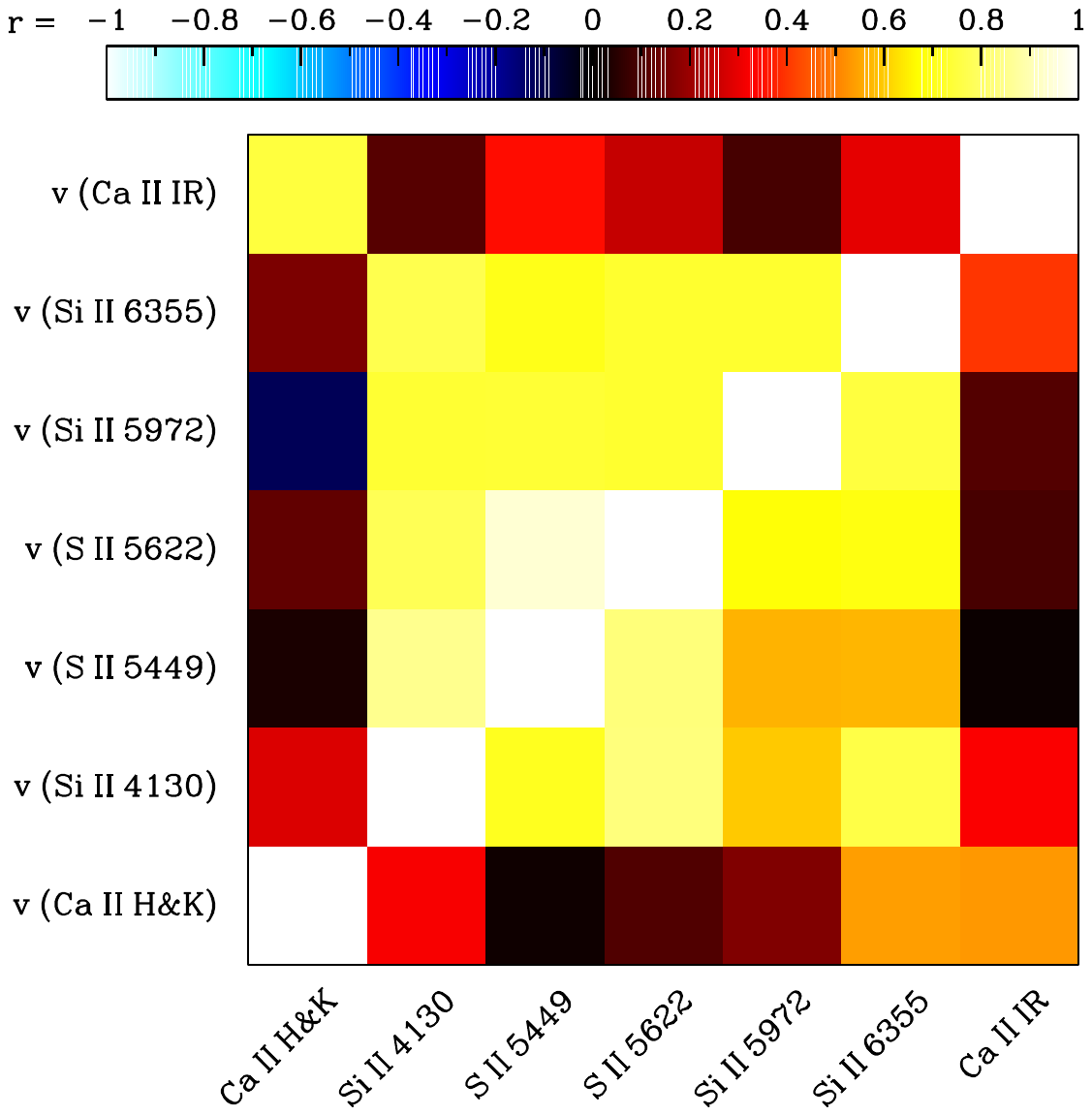}
 \caption{Correlation matrices for pairs of pseudo-equivalent widths (left) and expansion velocities (right) at maximum light for all the SNe~Ia analysed in this paper. The upper left off-diagonal triangle of each matrix shows results for the targeted sample, while the lower right off-diagonal triangle represents the SNe~Ia discovered in untargeted searches. As shown in the color scales on top of each panel, different colors correspond to different values of the Pearson coefficient ({\bf r}) with lighter colors indicating stronger correlation (or anti-correlation).
 \label{fig:correlation_matrices}}
\end{figure}

In order to provide a broader view of the possible correlations between the spectroscopic parameters under consideration,  we present in Figure~\ref{fig:correlation_matrices} correlation matrices for pairs of
pW values and expansion velocities at maximum light for the objects in our
targeted and untargeted samples, respectively.
\subsection {Correlations involving spectroscopic parameters
at maximum light and photometric properties}

\begin{figure}
\plottwo{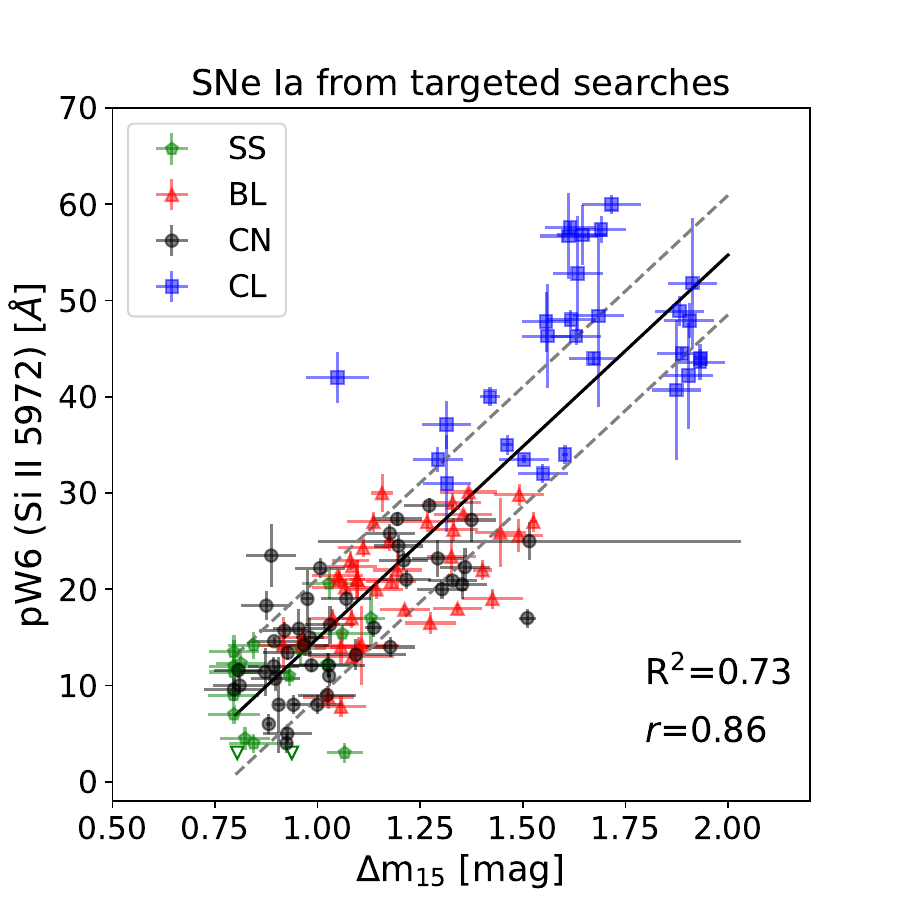}{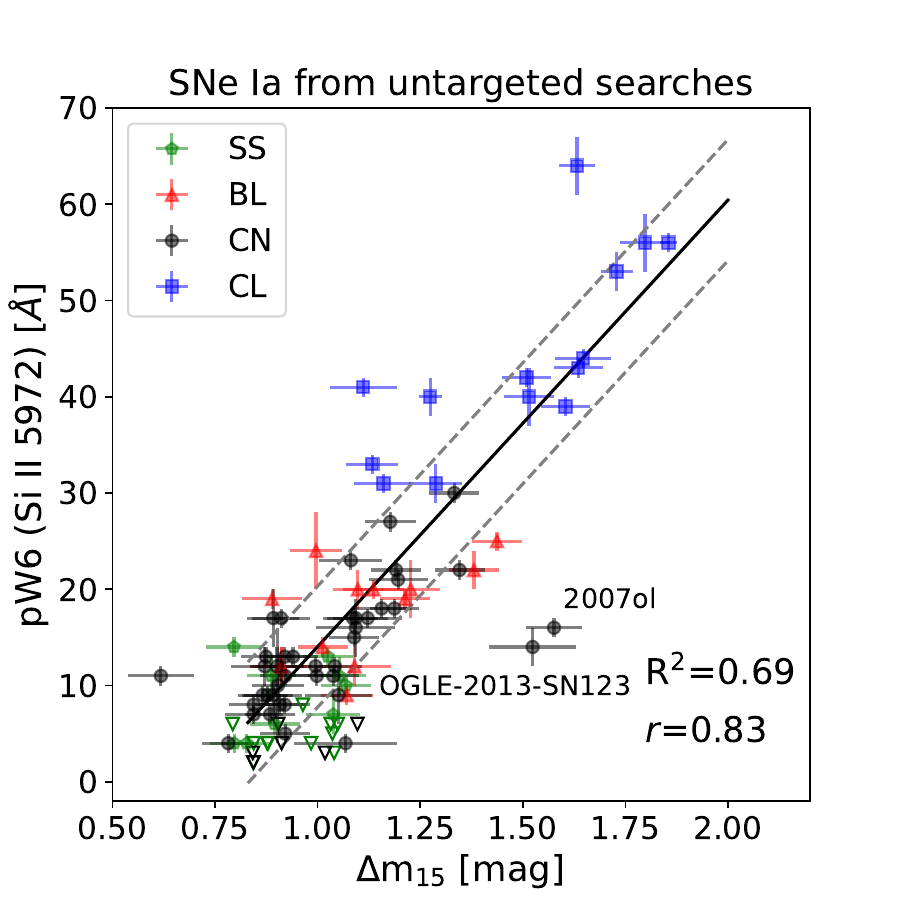}
\caption{Pseudo equivalent width at maximum light of \ion{Si}{2}~$\lambda$5972 versus
$\Delta{\rm m}_{15}$ 
for our two different samples of SNe~Ia discovered by targeted (left) and 
untargeted (right) searches.
The meaning of the symbols is as in Figure~\ref{fig:Branch}. The best-fit lines have slopes of 39.8$\pm$2.2~\AA ~per mag, 
and 46.4$\pm$3.5~\AA ~per mag, for the targeted and untargeted samples, respectively.
\label{fig:pW6_vs_DM15}}
\end{figure}

We briefly review here the strongest correlations between pW values at maximum
light and the light curve decline rate expressed by the typical $\Delta{\rm m}_{15}$
parameter, or the more recently defined color stretch parameter, $s_{BV}$.
Figure~\ref{fig:pW6_vs_DM15} shows the correlations between pW6 (\ion{Si}{2} 5972) and
$\Delta{\rm m}_{15}$.  That these two parameters correlate strongly was first pointed 
out by \citet{hachinger2006}.  The relations are similarly tight for both the targeted and untargeted samples.  
The most discrepant measurements in the untargeted sample correspond to the CNs SN~2007ol and SN OGLE-2013-SN-123.  
As explained in \S\ref{sec:notes}, the PESSTO
spectrum of the latter suffered from considerable host galaxy contamination.
While our attempt to correct for this problem was largely successful, the error
in the pW6 (\ion{Si}{2} 5972) measurement is likely underestimated due to
uncertainties in the subtraction of the host galaxy spectrum. Accounting for this extra source of uncertainty would bring the error
in pW6 up to $\pm$ 7 \AA. On the other hand, the spectrum of SN~2007ol looks good and does not show indications of host galaxy contamination.

\begin{figure}
\plottwo{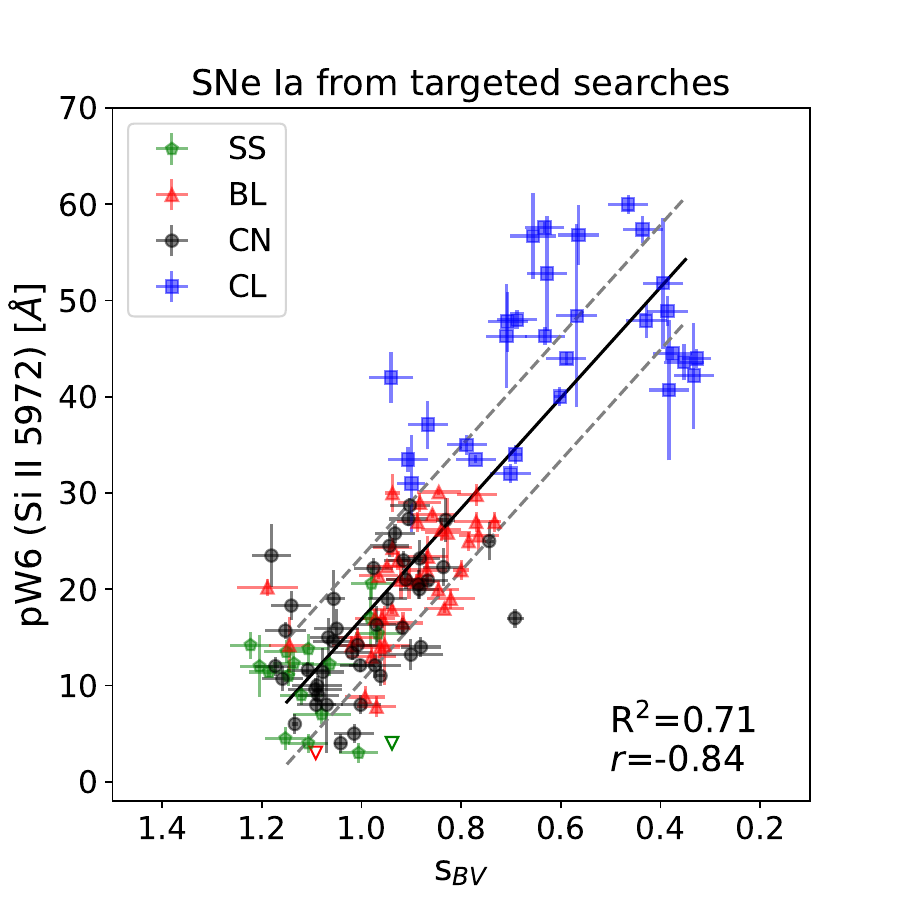}{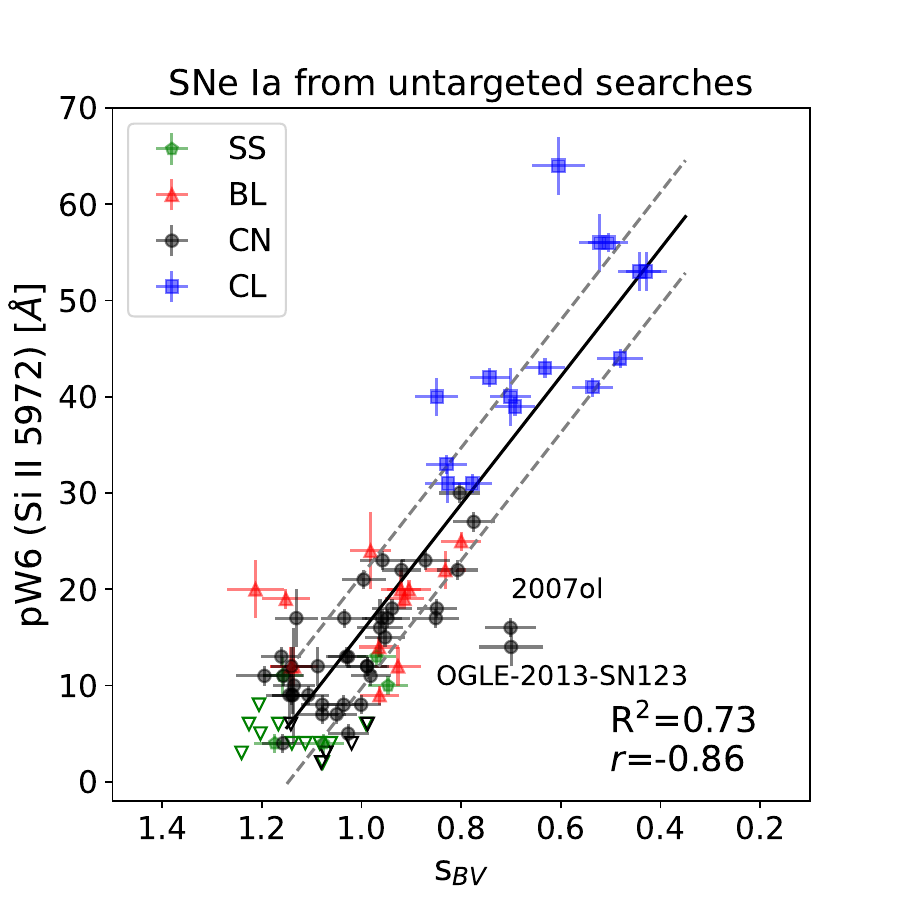}
\caption{Pseudo equivalent width at maximum light of \ion{Si}{2}~$\lambda$5972 versus
color stretch
for our two different samples of SNe~Ia discovered by targeted (left) and 
untargeted (right) searches.
The meaning of the symbols is as in Figure~\ref{fig:Branch}. The slopes of the
best-fit lines are  -57.4~\AA~$\pm$~3.2~\AA ~and -66.4~\AA~$\pm$~4.4~\AA ~for the
targeted and untargeted samples, respectively.
In this figure the abscissa has been inverted to facilitate comparison with Figure~\ref{fig:pW6_vs_DM15}.
\label{fig:pW6_vs_sBV}}
\end{figure}

Figure~\ref{fig:pW6_vs_sBV} displays the relationship between pW6 (\ion{Si}{2} 5972) and
$s_{BV}$.  We might have expected an improvement in the correlation since the
$s_{BV}$ parameter does a better job of discriminating between light curve shapes for
fast-declining events.  Nevertheless, the coefficients of determination and Pearson {\bf r} coefficients are similar to 
those obtained using $\Delta{\rm m}_{15}$.

\begin{figure}
\plottwo{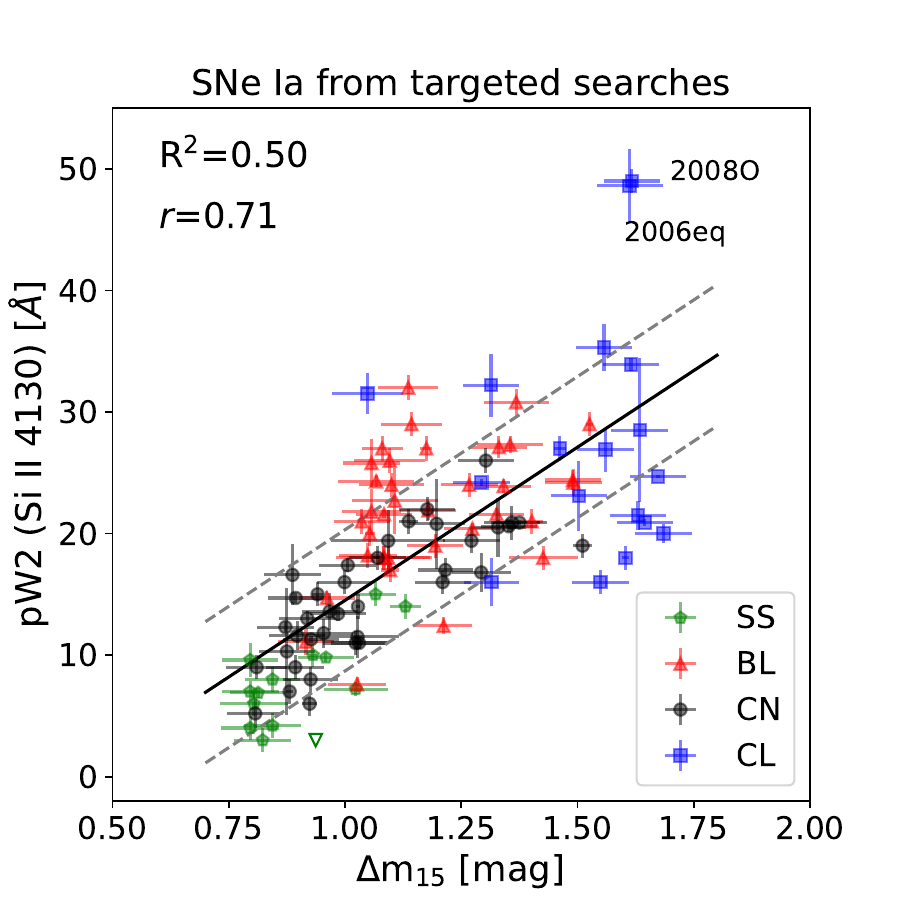}{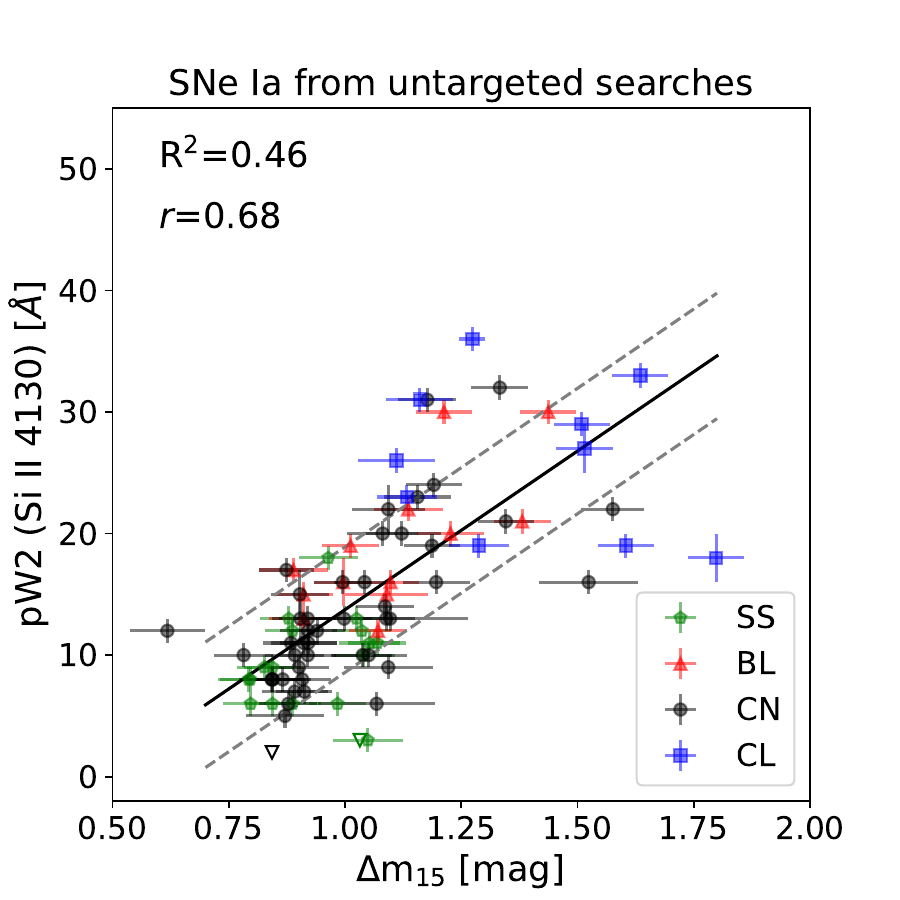}
\caption{Pseudo equivalent width at maximum light of \ion{Si}{2}~$\lambda$4130 versus
$\Delta{\rm m}_{15}$
for our two different samples of SNe~Ia discovered by targeted (left) and 
untargeted (right) searches.
The meaning of the symbols is as in Figure~\ref{fig:Branch}. The slopes of the best-fit lines are: 25.2$\pm$2.6~\AA ~per mag,  and 
26.1$\pm$3.1~\AA ~per mag, for the targeted and untargeted samples, respectively.
\label{fig:pW2_vs_DM15}}
\end{figure}

\begin{figure}
\plottwo{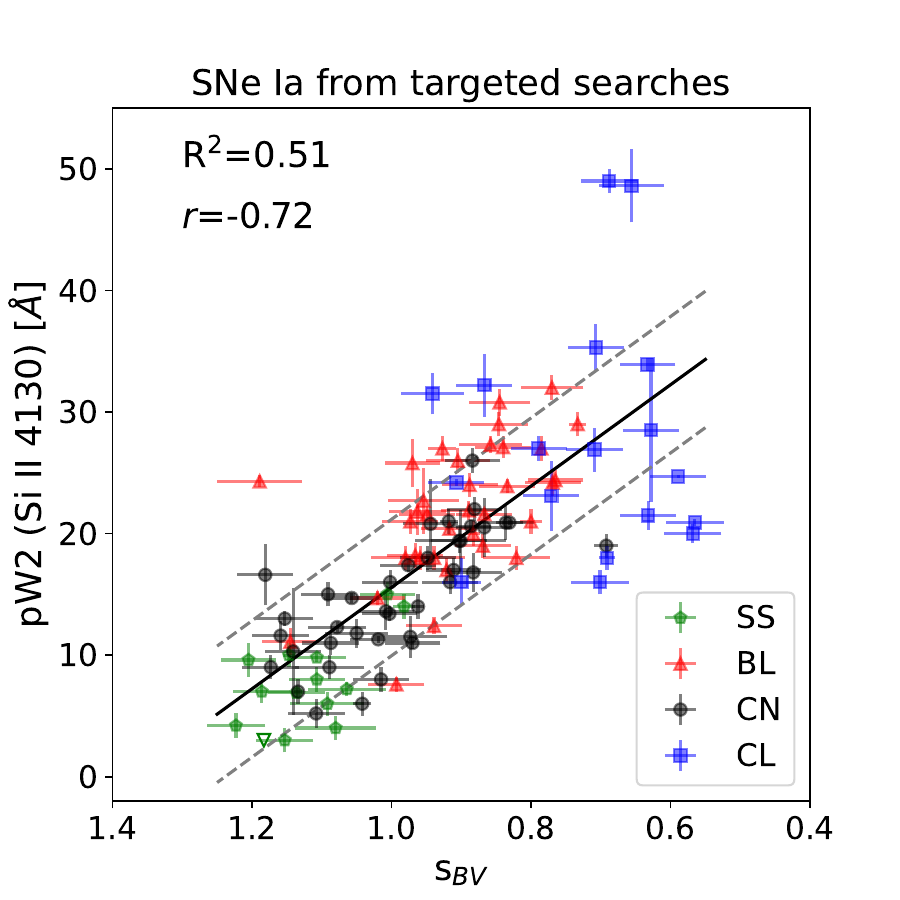}{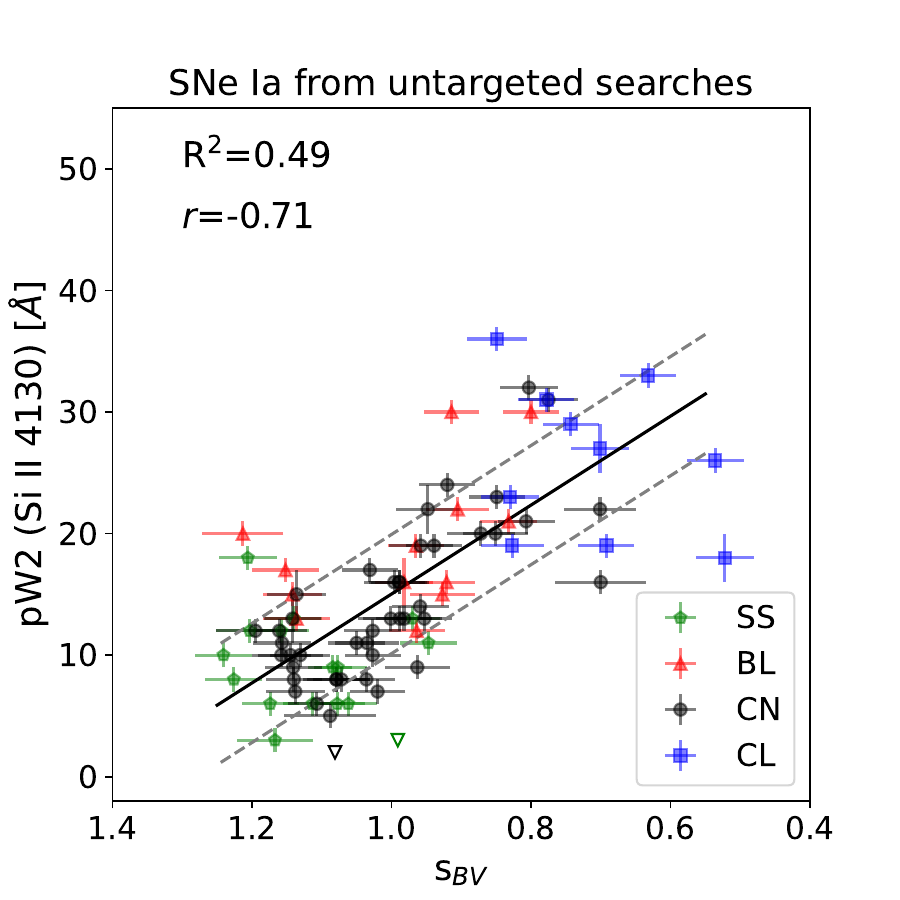}
\caption{Pseudo equivalent width at maximum light of \ion{Si}{2}~$\lambda$4130 versus
color stretch for our two different samples of SNe~Ia discovered by targeted (left) and 
untargeted (right) searches.
The meaning of the symbols is as in Figure~\ref{fig:Branch}. The slopes of the best-fit lines are -41.7~\AA ~$\pm$ 4.1~\AA ~and 36.6~\AA ~$\pm$ 4.0~\AA ~for the targeted and untargeted sample, respectively.
Note that the abscissa ($s_{BV}$) is inverted in this plot to facilitate comparison 
with Figure~\ref{fig:pW2_vs_DM15}.
\label{fig:pW2_vs_sBV}}
\end{figure}

\citet{folatelli2013} found that pW2 (\ion{Si}{2} 4130) also correlated strongly with
the light curve decline rate.  In Figures~\ref{fig:pW2_vs_DM15} and \ref{fig:pW2_vs_sBV},
this parameter is plotted against $\Delta{\rm m}_{15}$ and $s_{BV}$, respectively.  Again, 
it is somewhat surprising to see that usage of $s_{BV}$ does not significantly improve
the tightness of the correlations. This could be related to the fact that $s_{BV}$ seems not to work as
well as $\Delta{\rm m}_{15}$ for SS SNe Ia  (C. Burns, private communication).

\section {Summary}

In this paper, we have presented 230 optical spectra of 130 SNe~Ia observed during the course of the CSP-II
campaign, which was carried out between 2011-2015.  These data are complemented by an additional 148 optical
spectra of 30 SNe~Ia obtained during the CSP-I campaign (2004-2009) that were not included in the
paper by \citet{folatelli2013}.  Finally, we have appended to this paper an ``historical'' sample consisting of 53 spectra of 30 SNe~Ia observed by the
Cal\'an/Tololo Supernova Survey between 1990-1993, along with 163 additional spectra of 16 SNe~Ia obtained between 1986-2001, mostly
by members of the Cal\'an/Tololo team.  
A number of optical spectra of SNe Ia obtained in the course of the CSP campaigns already published in previous papers have also been considered in this work. 
Measurements of expansion velocities at maximum light, the \ion{Si}{2} $\lambda$6355
velocity decline parameter, $\Delta v_{20}$(\ion{Si}{2}), and pseudo-equivalent width features at maximum light in the system of \citet{garavini2007}
and \citet{folatelli2013} have been provided for as many of these SNe~Ia as possible.  These data have been combined
with measurements of the same parameters for the CSP-I SNe~Ia published by \citet{folatelli2013} to re-examine
the Branch diagram and a few of the strongest correlations of parameters found
for SNe~Ia discovered in targeted versus untargeted searches.  The most significant difference that we find is in the Branch
diagram for targeted searches, which contains proportionately more CL and BL objects than is the case for untargeted
searches.  This difference is ascribed to the fact that targeted searches are dominated by SNe~Ia discovered in luminous
galaxies, and that CL and BL events are known to preferentially occur in such galaxies.

\acknowledgments
The work of the CSP has been supported by the National Science Foundation
under grants AST0306969, AST0607438, AST1008343, AST1613426, AST1613455
and AST1613472. CSP-II also was supported in part by funding from the Danish Villum FONDEN (grant numbers 13261 and 28021) and the Independent Research Fund Denmark (IRFD) by a Sapere Aude II Fellowship awarded to M.D.S. Additional IRFD funding comes from Project 1  (8021-00170B)  and Project 2  (10.46540/2032-00022B) grants.
This paper includes data gathered with the 6.5 m Magellan Telescopes located at Las Campanas Observatory, the Gemini South, Cerro 
Pach\'on, Chile and Gemini North, Mauna Kea, Hawaii (Gemini Programs No. GS-2011B-Q-15-150-002 and GN-2013A-Q-68-82-002). Also based on observations collected at the European Organization for Astronomical Research in the Southern Hemisphere, Chile (ESO Programs 164.H-0376 and 0102.D-0095).
C.G. is supported by a research grant (25501) by the Villum FONDEN.
M.H. acknowledges support from FONDECYT-Chile through grants 92/0312 and 1060808; 
the National Science Foundation through grants GF-1002-96 and GF-1002-97;
the Association of Universities for Research in Astronomy, Inc., under NSF
Cooperative Agreement AST-8947990 and from Fundaci\'on Andes under project C-12984;
the Hubble Fellowship grant HST-HF-01139.01-A (awarded by the Space Telescope
Science Institute, which is operated by the Association of Universities for
Research in Astronomy, Inc., for NASA, under contract NAS 5-26555); and the Carnegie Postdoctoral Fellowship.
This work has been funded by ANID, Millennium Science Initiative, ICN12\_009.
L.G. acknowledges financial support from the Spanish Ministerio de Ciencia e Innovaci\'on (MCIN), the Agencia Estatal de Investigaci\'on (AEI) 10.13039/501100011033, and the European Social Fund (ESF)``Investing in your future'' under the 2019 Ram\'on y Cajal program RYC2019-027683-I and the PID2020-115253GA-I00 HOSTFLOWS project, from Centro Superior de Investigaciones Cient\'ificas (CSIC) under the PIE project 20215AT016, and the program Unidad de Excelencia Mar\'ia de Maeztu CEX2020-001058-M.
C.A., acknowledges support by NASA grants JWST-GO-02114, JWST-GO-02122 and JWST-GO-04436.024-A; and JPL grant SS03-17-23.
The research of J.C.W. and J.V. is supported by NSF AST-1813825. J.V. is also supported by
OTKA grant K-142534 of the National Research, Development and Innovation Office, Hungary.
The authors want to thank an anonymous referee for their kind report and useful suggestions.
\vspace{5mm}
\facilities{Magellan:Baade (IMACS imaging spectrograph), 
Magellan:Clay (LDSS3, MagE, MIKE), 
du~Pont (WFCCD, B\&C spectrograph, MODSPEC),
ESO:3.6m (EFOSC-2),
NTT (EMMI),
ESO:1.52m (B\&C spectrograph),
ESO: VLT (MUSE),
NOT (ALFOSC),
Gemini-South (GMOS),
Gemini:Gillett (GMOS),
CTIO:1.0m (B\&C spectrograph),
CTIO:1.5m (R-C spectrograph),
Blanco (R-C spectrograph),
UH:2.2m,
KPNO:2.1m (Gold camera),
MMT (Red channel),
Shane (Cassegrain spectrograph),
Nickel (Cassegrain spectrograph),
FLWO:1.5m (Z Machine),
La Silla-QUEST, CRTS, PTF, iPTF, OGLE, ASAS-SN, PS1, KISS, ISSP, MASTER, SMT,
Cal\'an/Tololo Supernova Survey.}

\bibliography{ms_refs}{}
\bibliographystyle{aasjournal}

\startlongtable
\begin{deluxetable}{lrcccclccc}
\tabletypesize{\footnotesize}
\tablecolumns{10}
\tablecaption {Summary of observations, spectroscopic and 
photometric properties of CSP SNe~Ia considered in this work
\label{CSP-prop}}
\tabletypesize{\scriptsize}
\tablehead{
\colhead{SN}&\colhead{N}&\colhead{Phase range [d]}&\colhead{SNID}&\colhead{Wang}&\colhead{Branch}&\colhead{z$_{Helio}$}
&\colhead{t(max) [MJD]}&\colhead{$\Delta{\rm m}_{15}$ [mag]}&\colhead{$s_{BV}$}\\
}
\colnumbers
\startdata
\sidehead {{\bf CSP I}}
\hline
2005gj&18&38.0, 341.4&Ia-CSM& \dots & \dots   & 0.0616 & 53657.30(1.28)$^a$&\dots & \dots\\
2007if$^b$&\dots &6.5, 30.5& Ia-pec (SC) & Normal& SS&0.0742&54348.4$^c$&\dots&\dots \\
2007jd&5&14.4, 44.3&Normal&HV&BL&0.0726&54361.9(1.0)&1.304(0.065)&0.891(0.054)\\
2007ol&7&$-$3.5, 20.1&Normal&Normal&CN&0.0559&54412.9(0.2)&1.576(0.068)&
0.701(0.051)\\
2007so&5&22.5, 41.0&Normal&\dots&\dots&0.0297&54428.0(0.5)&1.196(0.067)&0.925(0.041)\\
2007st&6&6.7, 35.9&Normal&\dots&\dots&0.0212&54454.4(0.7)&1.428(0.081)&0.863(0.046)\\
2008O&3&0.5, 28.4&Normal&HV&CL&0.0389&54491.8(0.6)&1.617(0.061)&0.688(0.041)\\
2008ae&4&0.6, 28.7&Ia-pec&\dots&\dots&0.0300&54508.63(0.60)$^d$&1.648(0.078)&\dots\\
\enddata

\tablenotetext{a}
{t(max) derived from $g$-band light curve \citep{prieto2007}}
\tablenotetext{b}
{03fg-like \citep{ashall2021}}
\tablenotetext{c}
{t(max) from \citep{scalzo2010}}
\tablenotetext{d} 
{t(max), $\Delta{\rm m}_{15}$ and $s_{BV}$ are derived from the observed light
 curves instead of \texttt{SNooPy} fits.}
\tablenotetext{e} {Observed during the course of the CSP, but not included in the final sample.}
\tablenotetext{f} {Phase estimated using SNID.}
\tablenotetext{g} {t(max) from \citep{marion2016}}
\tablenotetext{h} {t(max) from \citep{singh2018}}
\tablecomments{The meaning of the columns is as follows: 1: SN designation.
2: Number of spectra released in this work. 
Three dots mean that we are not presenting new spectroscopic
observations of the target, but, based on previously published spectra, we
include it in the discussion of SN properties at maximum light.
3: Phases (expressed in days) of the first and last spectra. 4: SNID classification.
5: Wang subtype. 6: Branch subtype. 7: Heliocentric redshift of the host. 
(*) The adopted redshift for LSQ12hnr is discussed in Section~\ref{sec:notes}
8: MJD of B maximum light from \texttt{SNooPy} fit when available. 
9: $\Delta{\rm m}_{15}$ (in magnitudes) from the \texttt{SNooPy} fit when available.
10: Color stretch $s_{BV}$ from \texttt{SNooPy} fit when available.
For data in columns 8, 9, and 10, errors are given in parentheses.}
\end{deluxetable}

\clearpage

\startlongtable
\begin{deluxetable}{lrcccclcccc}
\tabletypesize{\footnotesize}
\tablecolumns{11}
\tablecaption {Summary of observations, spectroscopic and photometric 
properties of historical SNe~Ia
\label{CT-prop}}
\tabletypesize{\scriptsize}
\tablehead{
\colhead{SN}&\colhead{N}&\colhead{Phase range [d]}&\colhead{SNID}&\colhead{Wang}&\colhead{Branch}&\colhead{z$_{Helio}$}
&\colhead{t(max) [MJD]}&\colhead{$\Delta{\rm m}_{15}$ [mag]}&\colhead{$s_{BV}$}&\colhead{Photometry}\\
}
\colnumbers
\startdata
1981B&\dots&\dots&Normal&HV    & BL    & 0.0060& 44671.5     &1.10(0.07)&\dots\\
1986G&17&$-$6.1, 55.8&91bg & 91bg  & CL    & 0.0018& 46561.192(0.077)  &1.420(0.025)&0.602(0.012) & 1\\
1989B&59&$-$8.2, 345.3&Normal& Normal&BL  & 0.0024& 47565.377(0.180)  &1.053(0.011)&0.883(0.010) & 2\\
1990N&7&$-$14.0, 37.6&Normal&Normal&SS   &0.0033 & 48082.208(0.040)  &0.931(0.009)& 1.147(0.008) & 3\\
1990O&6&$-$7.4, 19.6&Normal&Normal&CN & 0.0303 & 48075.945(0.171)  &0.905(0.012)&1.107(0.032) & 4\\
1990T$^a$&4&14.5, 34.7&Normal& \dots&\dots & 0.0404  &48083.068(0.210)  & 1.066(0.008)&0.985(0.017) & 4\\
1990Y$^a$&1&19      & Normal& \dots &\dots & 0.0360  &48114.129(0.612)  &1.035(0.023)&1.099(0.067) & 4\\
1990af$^a$&2&$-$3.6,$-$3.6& 91bg & 91bg&CN& 0.0506  &48195.867(0.120) & 1.516(0.021)&0.744(0.009) & 4\\
\enddata

\tablenotetext{a}
{Cal\'an/Tololo Supernova Survey.}
\tablenotetext{b}
{Phase estimated using SNID.}
\tablecomments{The meaning of the columns is as in Table\ref{CSP-prop}.
\texttt{SNooPy} fits are available for all SN other than 1993af and 1999cs.  Photometry references are:
(1) \citet{phillips1987};
(2) \citet{wells1994};
(3) \citet{lira98};
(4) \citet{hamuy1996};
(5) \citet{leibundgut1993}
(6) Suntzeff, N. B. (unpublished);
(7) \citet{strolger2002};
(8) \citet{lennarz2012};
(9) \citet{hamuy2002};
(10) \citet{krisciunas04b};
(11) \citet{krisciunas04a}.
}
\end{deluxetable}

\clearpage

\startlongtable
\begin{deluxetable}{lcrcccrrcc}
\tabletypesize{\footnotesize}
\tablecolumns{10}
\tablecaption{CSP Optical Spectra of SNe~Ia considered in this work
\label {CSP-spectra}}
\tablehead{
\colhead{Date [UT]}&\colhead{MJD}&\colhead{Phase [d]}&\colhead{Tel.}&
\colhead{Inst.}&\colhead{$\lambda$ range [\AA]}&\colhead{Res. [\AA]}&\colhead{exp. [s]}
&\colhead{airmass}&\colhead{rms }\\
}
\colnumbers
\startdata
\sidehead{\bf{CSP I}}
\sidehead{SN 2005gj}
\hline
2005-11-24.2$^{a}$   &   53698.2   &    38.0 &   DUP   &   MS   &   3780-7290   &   7.0   &   3x900   &   1.15   &   0.163  \\
2005-11-25.2   &   53699.2   &    39.0 &   DUP   &   MS   &   3780-7290   &   7.0   &   3x900   &   1.15   &   0.054  \\
2005-12-18.2   &   53722.2   &    60.7 &   NTT   &   EM   &   3400-5300   &   6.0   &   3x300   &   1.45   &   0.053   \\
2005-12-18.2   &   53722.2   &    60.7 &   NTT   &   EM   &   4000-10200   &   9.0   &   3x300   &   1.61   &   \dots   \\
2005-12-18.2   &   53722.2   &    60.7 &   NTT   &   EM   &   5800-10200   &   9.0   &   3x300   &   1.80   &   \dots   \\
2005-12-20.2   &   53724.2   &    62.5 &   DUP   &   WF   &   3800-9235   &   8.0   &   3x900   &   1.21   &   0.034   \\
2005-12-21.1   &   53725.1   &    63.4 &   DUP   &   WF   &   3800-9235   &   8.0   &   3x900   &   1.25   &   0.049   \\
2005-12-22.2   &   53726.2   &    64.4 &   DUP   &   WF   &   3800-9235   &   8.0   &   3x1200   &   1.32   &   0.040   \\
\enddata
\tablenotetext{a}{Spectrum corrected to match photometry.}
\tablenotetext{b}{Phase estimated using SNID.}
\tablenotetext{c}{Spectrum not corrected for telluric absorption.}
\tablecomments{Column 1: UT date of the observation. Column 2: Modified Julian Date.
Column 3: phase in days relative to time of maximum light corrected for redshift. 
Column 4: telescope code.
3P6: ESO 3.6~m Telescope at La Silla; BAA: Las Campanas Magellan I 6.5~m 
Baade Telescope; CLA: Las Campanas Magellan II 6.5~m Clay Telescope; 
DUP: Las Campanas 2.5~m du Pont Telescope; GEN: Gemini-N 8~m Telescope;
GES: Gemini-S 8~m Telescope;
NTT: ESO New Technology Telescope at La Silla; NOT: Nordic Optical Telescope at
Roque de los Muchachos, Canary Islands.   
Column 5: instrument code. 
AL: Alhambra Faint Object Spectrograph and Camera (ALFOSC);
BC: Boller \& Chivens spectrograph; EF: ESO Faint
Object Spectrograph and Camera (EFOSC-2); EM: ESO Multi-Mode Instrument (EMMI);
GM: Gemini Multiple-Object Spectrograph (GMOS); IM: Inamori Kyocera Magellan Areal
Camera and Spectrograph (IMACS); LD: Low Dispersion Survey Spectrograph (LDSS);
MA: Magellan Echellette spectrograph (MagE); 
MI: Magellan Inamori Kyocera Echelle spectrograph (MIKE);
MS: Las Campanas Modular Spectrograph (MODSPEC);
WF: Wide Field Reimaging CCD Camera (WFCCD). Column 6: observed spectral range expressed
in \AA. Column 7: spectral resolution in \AA; Column 8: exposure time in seconds.
Column 9: airmass at middle of exposure. Column 10: r.m.s. of the differences
between synthetic and observed broadband magnitudes (when available).
}
\end{deluxetable}

\clearpage

\startlongtable
\begin{deluxetable}{clrcllcc}
\tabletypesize{\footnotesize}
\tablecolumns{8}
\tablecaption{Optical Spectroscopic observations of historical SNe~Ia
\label{CT-spectra}}
\tablehead{
\colhead{Date [UT]}&\colhead{MJD}&\colhead{Phase [d]}&
\colhead{Tel.}&\colhead{Inst.}&\colhead{$\lambda$ range [\AA]}&\colhead{exp. [s]}&\colhead{airmass}\\ }
\colnumbers
\startdata
\sidehead{SN 1986G}
\hline  
1986-05-05.1&46555.1&$-$6.1&C60&RC60&4787-7179&\dots&\dots\\
1986-05-06.1&46556.1&$-$5.1&C60&RC60&4787-7179&\dots&\dots\\
1986-05-07.0&46557.0&$-$4.2&C60$,$BLA&RC60,RC4&3600-7154&\dots&\dots\\
1986-05-08.1&46558.1&$-$3.2&C60$,$BLA&RC60,RC4&3600-7162&\dots&\dots\\
1986-05-09.0&46559.0&$-$2.2&BLA&RC4&3600-5171&\dots&\dots\\
1986-05-10.2&46560.2&$-$1.0&BLA&RC4&3918-7018&\dots&\dots\\
1986-05-11.3&46561.3&0.1&BLA&RC4&3556-6638&\dots&\dots\\
1986-05-12.2&46562.2&1.0&BLA&RC4&3556-6638&\dots&\dots\\
\enddata
\end{deluxetable}

\startlongtable
\begin{deluxetable}{lrrrrrrrrr}
\tabletypesize{\scriptsize}
\tablecolumns{9}
\tablecaption {Pseudo equivalent widths at maximum light [\AA]
\label {pWs}}
\tablehead{
\colhead{SN}&\colhead {pW1}&\colhead{pW2}&\colhead{pW3}&\colhead{pW4}&
	\colhead{pW5}&\colhead{pW6}&\colhead{pW7}&\colhead{pW8}&\colhead{project}\\
\cline{2-10}
\colhead{}&\colhead{Ca {\sc ii} H\&K}&\colhead{Si {\sc ii} 4130}&
\colhead{Mg {\sc ii}}&\colhead{Fe {\sc ii}}&\colhead{S {\sc ii} W}
&\colhead{Si {\sc ii} 5972} &\colhead{Si {\sc ii} 6355} &\colhead{Ca {\sc ii} IR}
	&\colhead{}}
\startdata
\sidehead {\bf CN}
2007ol& 106(1)& 22(1)& 80(1)& 82(1)& 56(1)& 16(1)& 78(1)& \dots &CSP I\\
2008bz& 86(2)&  22(2)& 70(1) & 106(2)& 88(1)& 17(1)& 99(1)& 105(3)&CSP I\\
2008fr& 107(2)& 11(1)& 71(2)& 109(1)& 80(1)& 17(1)&82(2) & 72(5)&CSP I\\
2009I & 140(3)&9(1) &82(3)& 111(2) & 72(1) &10(1) &70(1) &89(3) &CSP I\\
2009cz& 120(1) & 9(1)& 92(1)& 131(1)& 67(1)& 12(1)& 81(1) &104(4)&CSP I\\
2009le& 60(2)& 11(1)& 104(2)& 124(1)& 62(1) & 9(1) & 86(1) & 114(2)&CSP I\\
ASASSN-14hr& 135(2)& 32(1)& 101(2)& 138(1)& 67(2)& 30(1)& 104(2)& 107(4)&CSP II\\ 
ASASSN-14hu& 150(2)& 11(1)& 97(2)& 115(1)& 76(1)& 7(1)& 83(1)& \dots&CSP II\\
\enddata
\tablecomments
{All pW values and their errors (in parentheses) are expressed in \AA.}
\end{deluxetable}

\clearpage

\startlongtable
\begin{deluxetable}{lrcccclclcr}
\tabletypesize{\scriptsize}
\tablecolumns{9}
\tablecaption {Expansion velocities at maximum light and Si {\sc ii} 6355 
velocity decline [km s$^{-1}$]
\label{vels}}
\tablehead{
\colhead{SN}&\colhead{Ca {\sc ii} H\&K}&\colhead{Si {\sc ii} 4130}&
\colhead{S {\sc ii} 5449}&\colhead{S {\sc ii} 5622}&\colhead{Si {\sc ii} 5972}
&\colhead{Si {\sc ii} 6355} &\colhead{Ca {\sc ii} IR}&
	\colhead{$\Delta{\rm v}_{20}$(Si)}&\colhead{project}\\
}
\startdata
\sidehead {\bf CN}
	2007ol&17110(212) &11217(98)& 9822(258)&9365(222)&11546(244)& 11577(195)&\dots &1559(331)&CSP I\\
2008bz&15632(266)& 10534(27)& 8883(58)& 8992(47)& 10704(72)&11303(39)&9710(257)&1253(151)&CSP I\\
2008fr&12296(141)&10175(95)& 9940(436)& 9208(117)& 10709(198)& 11098(118)& 12239(335)&\dots&CSP I\\
2009I & 12000(187)& 9764(84)& 9296(136)& 9696(94)& 9959(147)& 10768(79)& 12169(192)& 675(165)& CSP I\\
2009cz& 18359(65)& 8448(60)& 8270(136)& 8086(99)& 9560(192)& 9697(78)& 17545(248)&\dots&CSP I \\
2009le& 12570(247)& 12259(51)& 10495(145)& 10948(65)& 11071(181)& 12654(66)& 12273(396)&\dots&CSP I\\
	ASASSN-14hr& 17242(140)& 10650(70)& 8783(125)& 8527(137)& 11083(182)& 11747(306)& 11905(251)&\dots&CSP II \\
	ASASSN-14hu& 13054(191)& 10634(27)& 9797(63)& 10047(35)&\dots & 11511(36)&\dots & 1143(89)&CSP II\\
\enddata

\tablecomments{Velocities and their error estimates (in parentheses) 
are expressed in km~s$^{1}$. }
\end{deluxetable}

\clearpage

\begin{table}
\label{snake}
\centering
\caption {Temporal evolution of the \ion{Si}{2} $\lambda$6355 expansion velocities in the CSP I~\&~II sample. Phase: Rest frame phase of bin center; Exp.vel.: average expansion velocity; 
Sigma: standard deviation; N: number of data in bin.}
\begin{tabular}{rrrrrrrrrrrrrrrr}
\hline
        Phase & Exp. vel.& Sigma & N && Phase & Exp. vel.& Sigma & N &&
        Phase & Exp. vel.& Sigma & N  \\
\hline
        -9.5 & 12449 & 869 & 9&&1.5 &  10950 & 451 & 29&&
        13.5&  10261 & 600 & 5\\
        -7.5   & 11861 &  886   & 16&&3.5  &  10706 & 494 & 20&&
        15.5&  10319 & 665 & 6\\
        -5.5   & 11437 & 802& 17 &&5.5  &   10805 & 468 & 6&&
        19.0  &   9937 &   693 &   5\\
        -3.5&   11532 & 903 & 19&&7.5 &  10453 & 649 & 11&&
        21.0   &   9116 &   832 &   4\\
        -1.5 &  11066 & 696 & 26&&9.5 &  10190 & 653 & 12&&
        24.0  &   9812 &   543 &   6\\
        0   &  10861 &   619 &  71 &&11.5&  10290 & 514 & 8&&&&&\\
\end{tabular}
\end{table}
\end{document}